\documentclass[12pt,twoside]{article}
\usepackage{correl,breqn}
\def\TheTitle{ Green functions in the renormalized many-body perturbation theory}
\def\TheHeading{GF \& MBPT}
\def\TheAuthor{V\'aclav Jani\v s}
\def\TheInstitute{Institute of Physics,  The Czech Academy of Sciences,  Na Slovance 1999/2, CZ-182 21 Praha, Czech Republic}
\def\TheAddress{}
\def\TheChapter{5}                       % Chapter (will be assigned by us)
 \def\veck{\mathbf k}
\def\vecq{\mathbf q}

\newcommand{\be}{\begin{equation}}
\newcommand{\ee}{\end{equation}}
\newcommand{\bdm}{\begin{dmath}}
\newcommand{\edm}{\end{dmath}}

\newcommand{\Res}{\mathop{\rm Res}\nolimits}

\def\veck{\mathbf k}
\def\vecq{\mathbf q}

\def\vecQ{\mathbf Q}

\begin{document}
\MakeTitle           % displays title and table of contents
%%%%%%%%%%%%% put your text here:
\section{Introduction - Renormalization of the many-body perturbation theory}

The major objective of theoretical physics is to simplify the complexity of Nature  in that one breaks the whole into separate individual  phenomena or situations for which one identifies the relevant degrees of freedom that decide about the observed specific behavior. At the end, the theory provides models that should explain, mathematically formalize and quantify the observed phenomena. The relevance of each model is measured not only by its ability to reproduce the  experimental data. The strength of theoretical models lies in their predictive power of the behavior in yet unexplored situations. The basic idea behind simplified models with only a few degrees of freedom is that one will be able to solve them without further approximations. It is, however, a rare case and mostly only in specific, little realistic limiting situations.  That is why perturbation methods are ubiquitous in theoretical physics. 

Quantum dynamics arises from our inability to measure simultaneously the precise values of the fundamental variables. They are particle coordinate and velocity (momentum) in quantum mechanics. The observables are then represented by self-adjoint operators in a Hilbert space.  The operators  of not simultaneously measurable quantities do not commute. Lack of exact solutions of quantum models lies in the inability to find the exact spectrum of the generating Hamiltonian, being a sum of noncommuting operators. That is why the time-dependent perturbation theory was introduced in the one-particle quantum mechanics, in particular for the particle interacting with external fields. \cite{Shankar:1980aa}. 

Many-body quantum systems add to noncommutativity of the particle coordinates and momenta also indistinguishability of  identical particles and noncommutativity of the kinetic energy and the interaction in the Hamilton operator. This forces us to introduce the Fock Hilbert space and the occupation-number representation with creation and annihilation operators. Finding the exact dynamics of the quantum many-body models is generally beyond our reach. The only reliable technique to obtain quantitative results then is the perturbation theory with the many-body Green functions \cite{Abrikosov:1963aa}.  

Green functions were introduced in mathematics as a means to solve differential equations with boundary conditions \cite{Byron:1992aa}. The most attractive feature of Green functions are their analytic properties. The many-body Green functions are used in a slightly different context than in the solutions of differential equations. They solve exactly the unperturbed/noninteracting models with quadratic Hamiltonians, including boundary conditions, and enter as the fundamental ingredients in the perturbation expansion in powers of the non-quadratic particle interaction.  The importance of Green functions was emphasized early during the creation of the relativistic quantum field theory \cite{Tomonaga:1946aa,Schwinger:1949aa,Feynman:1949aa,Dyson:1949aa}. The way towards the application of the many-body Green functions also in nonrelativistic theories were set in Refs.~\cite{Martin:1959aa,Schwinger:1961aa} and extensively discussed in Ref.~\cite{Abrikosov:1963aa}. The important part of the application of the perturbation theory is its renormalization, inevitable in relativistic theories, being the only tool to extend the perturbation expansion from weak to strong coupling in many-body statistical models. The canonical way of performing renormalizations of the many-body perturbation theory systematically in terms of the one-particle Green function was proposed by Baym and Kadanoff \cite{Baym:1961aa,Baym:1962aa}.    

The Baym-Kadanoff approach addresses explicitly only the renormalization of the one-particle Green function. Explicit renormalizations of the two-particle Green function was introduced in Refs.~\cite{DeDominicis:1962aa,DeDominicis:1963aa,DeDominicis:1964aa,DeDominicis:1964ab}. The latter approach was later extended to the so-called parquet construction with an explicit two-particle self-consistency \cite{Jackson:1982aa,Bickers:1991aa,Bickers:1991ab}.  The one and two-particle Green functions cannot be renormalized independently. One has to obey two relations between them in order to keep the approximations conserving. One relation is established by the dynamical Schwinger-Dyson equation and the other is the Ward identity \cite{Baym:1962aa}.  The generic problem of the renormalized many-body perturbation theory is the inability to obey simultaneously the two relations between one and two-particle Green functions in approximate solutions \cite{Janis:2017aa}.    

The first level of the renormalization of the many-body perturbation theory are mean-field-like approximations with a set of static parameters to be determined self-consistently from an approximate thermodynamic potential. Such approximations are semiclassical with classical order parameters and Green functions are mostly not needed. The only way to include the true quantum dynamical effects into the renormalizations is to employ Green functions and to utilize their analytic properties to keep the approximations consistent and free of a spurious and unphysical behavior. I will review the way renormalizations of the many-body perturbation theory can be introduced in models of interacting fermions and show how to use the Green functions to derive consistent approximations, even having a mean-field character, in the strong-coupling regime.

\section{Generic models of interacting electrons \& quantum perturbations}

There are a few generic models of strongly interacting electrons that offer exact solutions in specific limits. They may  be used for testing the effectivity and reliability of the renormalized perturbation theory. The paradigm for strongly correlated electrons is the \index{Hubbard model}single-band Hubbard model the tight-binding Hamiltonian of which can be represented in second quantization as
\begin{align}\label{eq:Hubbard}
\widehat{H}_{H}&=\sum_{{\bf k},\sigma} \epsilon({\bf k})
   c^{\dagger}_{{\bf k}\sigma}
  c^{\phantom{\dagger}}_{{\bf k}\sigma}   +
  U\sum_{i}\widehat{n}_{i\uparrow}\widehat{n}_{ i \downarrow}  \,,
\end{align}
where $\epsilon(\veck)$ is the dispersion relation, $\widehat{n}_{ i\sigma}$ is the operator of the  density of particles with spin $\sigma$ on lattice site $\mathbf{R}_{i}$, and $U$ is the electron repulsion due to the screened Coulomb potential. This model can be solved exactly in the extreme limits of one and infinite dimensions. The exact solution in the former case can be reached at zero temperature via an algebraic Bethe ansatz \cite{Essler:1991aa}  while the full solution in the latter limit can be obtained only numerically \cite{Georges:1996aa}.     

There is a simpler version of the Hubbard model that offers an analytic solution in infinite dimensions. It is the \index{Falicov-Kimball model} spinless Falicov-Kimball model 
\begin{equation}\label{eq:FK}\widehat{H}_{FK}=-t\sum_{<ij>}c_{i}^{\dagger
}c_{j}+ \sum_i \epsilon_{i}f_i^{\dagger }f_i+ U\sum_ic_i^{\dagger
}c_i  f_i^{\dagger }f_i\,, 
\end{equation}
where $\epsilon_{i}$ are the atomic levels of the immobile electrons, and $<ij>$ indicates nearest neighbors. Unlike the Hubbard model, the Falicov-Kimball model does not describe a Fermi liquid that is why this model is rather limited in its applications and less suitable for testing the renormalized perturbation theory of correlated electrons \cite{Freericks:2003aa}.

The third  generic model of interacting electrons with an exact solution is the \index{single-impurity Anderson model} single-impurity Anderson model (SIAM) for the formation of the local magnetic moment. Its Hamiltonian reads 
\begin{equation}
\label{eq:SIAM}\widehat{H}_{SIAM}=-t\sum_{<ij>\sigma }c_{i\sigma }^{\dagger
}c_{j\sigma \,}+E_f\sum_\sigma f_\sigma ^{\dagger }f_\sigma +Uf_{\uparrow
}^{\dagger }f_{\uparrow }f_{\downarrow }^{\dagger }f_{\downarrow
}+\sum_{i,\sigma }\left( V_ic_{i\sigma }^{\dagger }f_\sigma +V_i^{*}f_\sigma
^{\dagger }c_{i\sigma }\right) \,, 
\end{equation}
where  $E_{f}$ is the impurity energy level and $V_{i}$ is the hybridization between the local impurity $f$ and the conduction $c_{i}$ electrons. Its solution serves as the input for the self-consistency equation of the Hubbard model in infinite dimensions. It offers a local model with strong dynamical quantum fluctuations leading at half filling to a quantum critical behavior and the strong coupling Kondo effect \cite{Tsvelick:1983aa,Hewson:1993aa}.  

All three Hamiltonians represent genuine many-body systems where the operator of the local particle interaction does not commute with the nonlocal operator of the kinetic energy, or the hybridization term of the SIAM.  Quantum fluctuations are important and must be appropriately treated in the perturbation theory. Only sufficient renormalizations can allow to continue reliably the perturbation expansion from weak to strong coupling.  Although we are interested in equilibrium properties of the many-body models we must introduce perturbations driving the systems out of equilibrium to explore all possible changes of the equilibrium state when crossing from weak to strong coupling. Most interestingly, the models may display a quantum critical behavior at low temperatures driven by cooperative quantum fluctuations. One has to choose the perturbing forces as general as possible not to miss any, sometimes unexpected, quantum phase transitions.

The fundamental quantity to be determined from the perturbation theory  is the grand potential $\Omega$ of the systems with appropriate external excitations represented by a time-dependent Hamiltonian $\widehat{H}_{ext}$. The perturbation theory treats the non-quadratic part of the Hamiltonian $\widehat{H}_{I}$ and the external nonequilibrium term $\widehat{H}_{ext}$ as a perturbation of the exactly solved quadratic part $\widehat{H}_{0}$, a set of harmonic oscillators. The generating functional of the perturbation expansion is        
\be
  \Omega[H_{ext}] =  - \beta^{-1} \log \text{Tr}\left[ \exp\left\{ -\beta \left(\widehat{H}_{0} - \mu \widehat{N} + \underbrace{\widehat{H}_{I} + \widehat{H}_{ext}}_{\mathrm{perturbation}} \right) \right\}\right] \,,
%\Omega_{ext} = -k_{B}T \ln \left[\Tr e^{-\beta \left(\widehat{H} + \widehat{H}_{ext}\right)} \right]
\ee
where $\widehat{N}$ is the number operator and $\beta = 1/k_{B} T$ is the inverse temperature. The external part of the total perturbation is quadratic but it generally contains an arbitrary combination of the creation and annihilation operators. There are four general quadratic terms for spin one-half fermions.  We denote the nonlocal external field perturbing the densities of  particles with spin $\sigma$ by  $\eta^{||}_\sigma(1,2)$. We used a short-hand notation for the space-time and spin variables  $l = (\mathbf{R}_l,\tau_l,\sigma_l)$. This field generates longitudinal spin fluctuations. Analogously we denote $\eta^{\perp}(1,2)$ the field that generates transversal spin fluctuations. We further introduce fields $\xi^{\perp}(1,2)$ and $\xi^{||}_\sigma(1,2)$ generating transversal and longitudinal charge fluctuations, respectively. Fields  $\eta^{\perp}$, $\xi^{\perp}$, and $\xi^{||}_\sigma$ are complex except for  $\eta^{||}_\sigma$. The complex character of the excitation field indicates that the perturbation breaks either spin, charge or both. Complex fields are not measurable but when the linear response to them is broken they generate a quantum long-range order.  The most \index{general quantum perturbation} general quantum perturbation can then be represented as 
\begin{multline}
  \label{eq:H-ext}
  \widehat{H}_{ext}=\int
      d1d2\left\{\sum_\sigma \eta^{||}_\sigma(1,2) c^{\dagger}
      _{\sigma}(1) c^{\phantom{\dagger}}_{\sigma}(2) \quad \ldots \text{conserves charge \& spin}\right. \\
  \left.\!\!\! +\left[\eta^{\perp}(1,2)c^{\dagger}_{\uparrow}(1)c^{\phantom{\dagger}}
      _{\downarrow}(2) +\bar{\eta}^{\perp}(1,2)c^{\dagger}_{\downarrow}(2)
      c^{\phantom{\dagger}}_{\uparrow}(1)\right]\quad\ldots  \text{conserves charge} \right. \\ 
     \left. +\ \left[\bar{\xi}^{\perp}
      (1,2)c^{\phantom{\dagger}}_{\uparrow}(1)c^{\phantom{\dagger}}
      _\downarrow(2)+\xi^{\perp}(1,2)c^{\dagger}_{\downarrow}(2)
      c^{\dagger}_{\uparrow}(1)\right]     \quad\ldots \text{conserves spin}\right. \\   
 \hspace*{-22pt} \left.   + \sum_{\sigma} \left[\bar{\xi}^{||}_\sigma
      (1,2)c^{\phantom{\dagger}}_{\sigma}(1)c^{\phantom{\dagger}}_{\sigma}
      (2) +\xi^{||}_\sigma(1,2) c^{\dagger}_{\sigma}(1)c^{\dagger}_{\sigma}(2)\right] \quad\ldots  \text{changes charge \& spin}\right\} \,,
\end{multline}
where the bar denotes complex conjugation. The possible long-range orders generated by this perturbation with fields  $\eta^{||}_\sigma$, $\eta^{\perp}$,  are longitudinal and transversal magnetic order and singlet and triplet superconductivity for fields $\xi^{\perp}$, and $\xi^{||}_\sigma$, respectively \cite{Janis:1999aa}.

\section{Static renormalizations}
\label{sec:StaticRen}

The exactly solvable statistical models are mostly those with quadratic Hamiltonians. The perturbation theory uses them as unperturbed models and expands thermodynamic quantities  in powers of the non-quadratic interacting Hamiltonians. The first step to improve upon such a bare perturbation expansion is to introduce a renormalization of the parameters characterizing the unperturbed system. This is the idea of mean-field theories.  This static renormalization optimizes the initial state of the perturbation expansion.

\subsection{Variational mean-field theories}
\label{sec:MFT}

There are several ways how to derive mean-field approximations. The most advanced construction of comprehensive mean-field theories  is presently the limit to infinite spatial dimensions of a hypercubic lattice with an appropriate rescaling of the off-diagonal (nonlocal) elements in the model Hamiltonian \cite{Thompson:1974aa}. This construction leads to a  dynamical mean-field theory (DMFT) of quantum many-body models \cite{Georges:1996aa}. This limit is not analytically solvable for the Hubbard model and cannot hence be used as a renormalized starting point of the perturbation theory. The original idea behind the classical mean-field approximations is to create optimal upper or lower bounds on the equilibrium thermodynamic potential by renormalizing the parameters of the exactly solvable unperturbed Hamiltonian.     

We split the total Hamiltonian into an exactly solvable part $\widehat{H}_{0}$ and a correction $\Delta\widehat{H}$
\be
\widehat{H} = \widehat{H}_{0} + \Delta\widehat{H} 
\ee
and use the \index{Gibbs-Bogoljubov inequality} Gibbs-Bogoljubov inequality for the grand potential 
\be
\Omega\left\{\widehat{H} \right\} \le \Omega\left\{\widehat{H}_{0} \right\} + \left\langle  \Delta\widehat{H}\right\rangle_{0} \,.
\ee
Here $\langle  \Delta\widehat{H}\rangle_{0}$ is the statistical average of the correcting term in the unperturbed system. A mean-field approximation is obtained by optimizing the upper bound with spin-dependent energies $E_{\sigma}$ renormalizing the chemical potential $\mu$ and the external magnetic field $h$. That is, we choose 
 $\widehat{H}_{0}= \sum_{{\bf k},\sigma} \epsilon({\bf k})
   c^{\dagger}_{{\bf k}\sigma}
  c^{\phantom{\dagger}}_{{\bf k}\sigma} +  \sum_{\mathbf{i}\sigma}\left( E_{\sigma} - \mu - \sigma h\right)\widehat{n}_{{\bf i}\sigma}$ 
and the correction $\langle  \Delta\widehat{H}\rangle_{0}/N = Un_{\uparrow}n_{\downarrow} - \sum_{\sigma}E_{\sigma}n_{\sigma}$, where $n_{\sigma}= N^{-1}\sum_{\mathbf{i}}\langle \widehat{n}_{\mathbf{i}\sigma}\rangle_{0}$. The lowest upper bound is then produced by the Hartree mean-field solution $E_{\sigma}= Un_{-\sigma}$.

There is also a variational way to construct a lower bound on the equilibrium grand potential \cite{Janis:1993aa}. We decompose the total Hamiltonian by using positive  normalized parameters  $\lambda_{\alpha}\ge 0$ and  $\sum_{\alpha}\lambda_{\alpha}=1$,
\be
\widehat{H} = \sum_{\alpha}\lambda_{\alpha}\widehat{H}_{\alpha} \,.
\ee
Using convexity property of the grand potential we obtain an inequality
\be
\sum_{\alpha}\lambda_{\alpha}\Omega\left\{\widehat{H}_{\alpha} \right\} \le  \Omega\left\{\widehat{H} \right\} \,.
\ee
The grand potential  of the Falicov-Kimball model from Eq.~\eqref{eq:FK} can be expressed via  analytic functions exactly in infinite spatial dimensions \cite{Janis:1991aa}.  We can then use the following decomposition of the Hubbard 	Hamiltonian with two parameters $\lambda_{\alpha}$, $\alpha= 1,2$, 
$\lambda_{\alpha}\widehat{H}_{\alpha}= \sum_{{\bf k}} \epsilon({\bf k})
   c^{\dagger}_{{\bf k}\alpha}
  c^{\phantom{\dagger}}_{{\bf k}\alpha} + \lambda_{\alpha}\sum_{{\bf i}\sigma}\left( E_{\sigma} - \mu_{\sigma} \right)\widehat{n}_{{\bf i}\sigma} + U\lambda_{\alpha}\sum_{{\bf i}}\widehat{n}_{{\bf i}\uparrow}\widehat{n}_{{\bf i}\downarrow} 
  $. The grand potential of the component Hamiltonians is known analytically in the DMFT. We can hence maximize the lower bound on the Hubbard model by finding the optimal decomposition parameters $\lambda_{\alpha}$. The derived mean-field approximation of the maximal lower bound is a thermodynamically consistent extension of the  Hubbard III approximation \cite{Janis:1993ab}.   

%\subsection{Hubbard-Stratonovic transformation}

\subsection{Fermi liquid}

The static mean-field approximations are short of an important feature of correlated electrons in metals.  It is the concept of quasiparticles with a finite lifetime. They are generated only by dynamical correlations of the low-lying excitations of the ground state. The quasiparticles were introduced by Landau in his theory of a Fermi liquid \cite{Landau:1957aa}. Fermi liquid was initially presented as a phenomenological theory but it has a profound impact on all theories of interacting fermions. It qualitatively correctly describes  the behavior of the low-lying fermionic excitations if the actual interaction can be reached adiabatically from zero (Fermi gas) without coming across a phase transition or, equivalently, a divergence in the perturbation theory. Consequently, the Fermi-liquid theory reduces the impact of the particle interaction only to a  renormalization of the parameters of the Fermi gas.       

The fundamental assumption of the Landau Fermi-liquid theory is the existence of an energy functional depending only on squares of the instantaneous densities of the low-lying excitations $\delta n_{\veck,\sigma}(t)$ caused by a general weak external potential $V_{\veck\sigma}$. The macroscopic generated averaged energy functional is 
\be
E\left[\delta n_{\veck,\sigma}\right] = \sum_{\veck,\sigma}\left(\epsilon(\veck) + V_{\veck\sigma}\right)\overline{\delta n_{\veck,\sigma}(t)} + \frac 1{2V}\sum_{\veck\veck',\sigma\sigma'}f(\veck,\veck';\sigma,\sigma')\overline{\delta n_{\veck,\sigma}(t)\delta n_{\veck',\sigma'}(t)}\,,
\ee
where the bar denotes averaging over the microscopic time scales $t$. The bare particle interaction of the model Hamiltonian was replaces  by the \index{Landau scattering function} Landau scattering function $f(\veck,\veck';\sigma,\sigma')$ standing for the screened dynamical interaction of quasiparticles, densities of the low-lying excitations. It is assumed as the input in the phenomenological Fermi liquid. 

Due to the ergodic theorem the time averaging can be replaced by a statistical one. And since the energy functional should depend maximally on squares of the densities of the excited states, the Hatree decoupling for the product of time-dependent densities of the excitations holds 
\be
\overline{\delta n_{\veck,\sigma}(t)\delta n_{\veck',\sigma'}(t)} = \left\langle\delta n_{\veck,\sigma}\right\rangle\left\langle\delta n_{\veck',\sigma'}\right\rangle \,.
\ee
The elementary excitations correspond to the eigenstates of the Fermi gas and hence the statistical average has the fermionic form 
\begin{align}
 \left\langle\delta n_{\veck,\sigma}\right\rangle &= \frac 1{\exp\left\{-\beta\left(\epsilon_{\veck}+ V_{\veck\sigma} + U_{\veck,\sigma}\right)\right\} + 1} - \theta\left(k_{F} - k\right) \,, 
\end{align}
with the Fermi momentum $k_{F}$ and an effective potential 
\begin{align} 
U_{\veck,\sigma} &= \frac 1{V}\sum_{\veck',\sigma'}f(\veck,\veck';\sigma,\sigma')\left\langle\delta n_{\veck',\sigma'}\right\rangle \,.
\end{align}
The statistical averaging was chosen so that the density of the particle-like excitations ($k>k_{F}$) is positive while the density of the hole-like excitations ($k<k_{F}$) is negative. The low-lying excitations are affected only by the Landau scattering function at the Fermi energy $E_{F}$  and $|k| \approx k_{F}$. That is, the only relevant physical scale in this theory is the Fermi energy. The Landau scattering function can either be determined from experimental data or can be calculated from a microscopic theory. At the end, the Fermi liquid is a Fermi gas with renormalized thermodynamic quantities \cite{Baym:1991aa}. In this sense it is a semi-classical theory with only averaged quantum dynamical fluctuations.

\section{Dynamical corrections and Green functions}

So far we introduced only static renormalizations of the equilibrium thermodynamic quantities. The genuine quantum fluctuations due to noncommutativity of the kinetic-energy and the particle-interaction operators can be obtained only from a fully dynamical perturbation theory. The only way to keep all the necessary information from the quantum fluctuations is to use the Matsubara formalism and the many-body Green functions.   

\subsection{Green functions, Matsubara formalism \& analytic continuation}

The dispersion relation between energy and momentum from the model Hamiltonian is generally obeyed only at the energy (mass) shell. Since we are unable to find the spectrum
of the full Hamiltonian we have to go off the mass shell and to introduce 
time-dependent excitations for which the energy is detached from the momentum. We introduce an imaginary time $\tau\in(0,\beta)$ and let propagate the creation and annihilation operators with the
unperturbed free-electron Hamiltonian $\widehat{H}_0 = \sum_{{\bf k}\sigma}
\left(\epsilon({\bf k}) - \mu - \sigma h\right) c^{\dagger}_{{\bf k}\sigma}
c^{\phantom{\dagger}}_{{\bf k}\sigma}$, where $\mu$ is the chemical potential and $h$ is the external magnetic field. We define
$c^{\phantom{\dagger}}_{{\bf k}\sigma}(\tau) = \exp\{\tau
\widehat{H}_0\} c^{\phantom{\dagger}}_{{\bf k}\sigma} \exp\{-\tau
\widehat{H}_0\}$ and analogously the time-dependent hermitian-conjugate creation operator, $c^{\dagger}_{{\bf k}\sigma}(\tau) = \exp\{\tau
\widehat{H}_0\} c^{\dagger}_{{\bf k}\sigma} \exp\{-\tau
\widehat{H}_0\}$. Notice that $c^{\dagger}_{{\bf k}\sigma}(\tau) \neq c_{{\bf k}\sigma}(\tau)^{\dagger}= c^{\dagger}_{{\bf k}\sigma}(-\tau)$.  Moving the creation and annihilation operators off the mass shell we
can introduce Green functions as time-ordered moments of the
density-matrix operator
\begin{dmath}\label{eq:Omega-moments-time}
G_{(n)}(1,\ldots, n, \bar{n},\ldots,\bar{1}) = \frac{(-1)^{n}}{\hbar^{n}}\left\langle \frac
1{\mathcal{Z}} \text{Tr}_0\mathcal{T}\left[c^{\phantom{\dagger}}(1)\ldots
c^{\phantom{\dagger}}(n), c^\dagger(\bar{n}) \ldots c^\dagger(\bar{1})
\exp\left\{-\int_0^\beta d\tau \widehat{H}_I(\tau)
\right\} \right]\right\rangle_{av}
\end{dmath}
%\end{subequations}
where we denoted $\text{Tr}_0 \widehat{X} = \text{Tr}\left[\widehat{X}
\exp\{-\beta(\widehat{H}_0\}\right]$ and $\mathcal{Z} =
\text{Tr}_0 \mathcal{T}\exp\left\{-\int_0^\beta d\tau\widehat{
H}_I(\tau)\right\}$ is the partition sum.  

The \index{Matsubara formalism} Matsubara formalism \cite{Matsubara:1955aa} uses the grand potential with a fixed chemical potential during the perturbation expansion of the functional of the partition sum. The propagation in the imaginary time can then be diagonalized by a Fourier series of the periodic extensions of interval $(-\beta,\beta)$ with imaginary even, bosonic $i\nu_{m}= 2m\pi i k_{B}T$, and odd, fermionic $i\omega_{n}= (2n + 1)\pi ik_{B}T$, frequencies where $n,m$ are  integers. 
 
We can represent the partition function, the generator of the perturbation expansion,  via a functional integral over the  Grassman anticommuting variables  $\psi$ and $\psi^{*}$ replacing the time ordering of the operators in the interaction picture by an unperturbed (inverse) Green function  diagonal in momenta and Matsubara frequencies  
\bdm
\mathcal{Z} = \int \mathcal{D}\psi \mathcal{D}\psi^* \exp\left\{\sum_{\mathbf{k}}\sum_{n\sigma}e^{i\omega_{n}0^{+}}
\psi^*_{n\sigma}(\mathbf{k}) (i\omega_n  + \mu - \epsilon(\mathbf{k}))\psi_{n\sigma}(\mathbf{k})  - U
\sum_{i}\int_0^\beta d\tau\ \widehat{n}^d_\uparrow(\tau, \mathbf{R}_{i})
\widehat{n}^d_\downarrow(\tau, \mathbf{R}_{i})\right\}\,.
\edm
The local interaction term is diagonal in the imaginary time and does not commute with the unperturbed, kinetic part, diagonal in momenta and Matsubara frequencies. The expansion of the partition sum in the powers of the interaction strength is the  gist of the many-body perturbation theory.   

The problem of the sums over the Matsubara frequencies is that one must sometimes use the normalization factor $e^{i\omega_{n}0^{+}}$ defining the proper limit to large frequency values. This can be avoided by analytic continuation  from the discrete set of  Matsubara frequencies on the imaginary axis to continuous real frequencies weighted by the appropriate statistical distribution. The sums over the fermionic and bosonic frequencies can be analytically continued to \index{spectral integrals} spectral integrals as follows
\begin{subequations}
\begin{align}
\frac{1}{\beta}\sum_n F(i\omega_n)e^{i\omega_{n}0^{+}} &\rightarrow\  - \int_{-\infty}^{\infty} \frac{d\omega}{\pi} f(\omega)\Im F(\omega+i0)  + \sum_i f(\zeta_i)\Res[F,\zeta_i]\,,
\\
\frac{1}{\beta}\sum_m F(i\nu_m)e^{i\omega_{m}0^{+}} &\rightarrow \ P\int_{-\infty}^{\infty} \frac{d\omega}{\pi} b(\omega)\Im F(\omega+i0) -\sum_i b(\zeta_i)\Res[F,\zeta_i] \,,
\end{align}
\end{subequations}
where $P$ stands for the principal value of the integral and we denoted $f(x) = 1/(e^{\beta x} + 1)$ and $b(x) = 1/(e^{\beta x} - 1)$ the Fermi and Bose distribution functions, respectively. Further on,  $\zeta_{i}$ are poles of complex function $F(z)$ outside  the real axis. One can uniquely analytically continue from the set of the values at the Matsubara frequencies only if infinity is a point of analyticity of the continued function.

\subsection{One-particle Green function - Schwinger and Dyson equations}
%\label{sec:GF}

The basic ingredient of the perturbation theory is the one-particle Green function. Its operator definition in the \index{Heisenberg picture} Heisenberg picture without splitting the full Hamiltonian into the unperturbed part and interaction reads  
\be\label{eq:GF-operator}
\widehat{G}(\tau,\sigma;\bar{\tau},\bar{\sigma}) = - \frac 1\hbar \text{Tr}\left\{\widehat{\rho}_{H}\ \mathcal{T}\left[ \widehat{\psi}_{\sigma}^{\phantom{\dagger}} (\tau)
\widehat{\psi}(\bar{\tau})_{\bar{\sigma}}^{\dagger}\right]\right\} \,.
\ee
We introduced the full density matrix $\widehat{\rho}_{H} = \exp\left\{ -\beta\widehat{H}\right\}/  \text{Tr}\exp\left\{ -\beta\widehat{H}\right\}$. We denoted $\widehat{\psi}$ and $\widehat{\psi}^{\dagger}$ the field operator and its hermitian conjugate. The time evolution is controlled by the full Hamiltonian, that is $\widehat{\psi}_{\sigma}(\tau) = \exp\{\tau\widehat{H}\} \widehat{\psi}_{\sigma} \exp\{-\tau\widehat{H}\}$.  The time-dependent Green function from Eq.~\eqref{eq:GF-operator} is an operator (matrix) in the spatial coordinates.  The number Green function then is $G(1,\bar{1}) = \langle \mathbf{R}_{1}|\widehat{G}(\tau_{1},\sigma_{1};\tau_{\bar{1}},\sigma_{\bar{1}})|\mathbf{R}_{\bar{1}}\rangle$. 

We cannot directly evaluate the Green functions in the Heisenberg picture since the full Hamiltonian is not diagonalizable. We separate the quadratic part of the total Hamiltonian from the non-quadratic one and come over to the Dirac interaction picture with the time propagation controlled by the unperturbed Hamiltonian as in Eq.~\eqref{eq:Omega-moments-time}. The impact of  the particle interaction $U$ on the one-particle Green function can be expressed via a \index{Schwinger integral equation} Schwinger integral equation  \cite{Baym:1961aa}
\be
\int d\bar{1}G^{(0)-1}(1,\bar{1})G(\bar{1},1^\prime) = \delta(1 - 1^\prime) -
 \int d\bar{1} U(1 - \bar{1}) G_{(2)}(1\bar{1}^-,1^\prime\bar{1}^+) \,,
\ee
where we introduced a  two-particle Green function
\be
G_{(2)}(1\bar{1},3\bar{3}) =  \frac 1 {\hbar^2}\text{Tr}\left\{\widehat{\rho}\
\mathcal{T}\left[\widehat{\psi}_{\sigma_{1}}(\mathbf{R}_{1},\tau_{1})\widehat{\psi}_{\sigma_{3}}(\mathbf{R}_{3},\tau_{3})\widehat{\psi}_{\sigma_{\bar{3}}}(\mathbf{R}_{\bar{3}},\tau_{\bar{3}})^\dagger
\widehat{\psi}_{\sigma_{\bar{1}}}(\mathbf{R}_{\bar{1}},\tau_{\bar{1}})^\dagger \right]\right\} \,.
\ee
The superscripts $\pm$ indicate whether we approach the equal-time limit of two time variables from above or below. 
The unperturbed Green function has an explicit algebraic expression in the Matsubara formalism
\be
G^{(0)}_{\sigma}(\veck,i\omega_{n}) = \frac 1{i\omega_{n} + \mu + \sigma h - \epsilon(\veck)} \,.
\ee

The Schwinger equation does not determine the one-particle Green function in closed form. One also must know the impact of the interaction on the two-particle Green function. If we go on with the two-particle Green function as in the Schwinger equation, we then obtain a relation between the two-particle and a three-particle Green function. It is generally not an effective way to evaluate Green functions.  It is better to formally close the dynamical equations for the Green functions in the same particle space by introducing irreducible functions. The one-particle irreducible function is a self-energy $\Sigma(\bar{1},2)$. The self-energy is used to  close the equation for the one-particle Green function within the one-particle representation space via a \index{Dyson equation} Dyson equation     
\begin{equation}
  G(1,\bar{1}) =  G^{(0)}(1 - \bar{1}) +  \sum_{3,\bar{3}}G^{(0)}(1 - \bar{3}) \Sigma(\bar{3},3) G(3,\bar{1}) \,.
\end{equation}  
Since the self-energy has the same dependence on the internal variables as the Green function, the Dyson equation becomes algebraic in momentum and frequency representation in equilibrium. The full equilibrium Green function in the Matsubara formalism then is
\be
G_{\sigma}(\veck,i\omega_{n}) = \frac 1{i\omega_{n} + \mu + \sigma h - \epsilon(\veck) - \Sigma_{\sigma}(\veck,i\omega_{n})} \,.
\ee
The Dyson equation is only a formal relation determining the Green function from the self-energy. The latter must be obtained from the perturbation theory.

\subsection{Two-particle vertex - Bethe-Salpeter equations}

\begin{figure}
  \begin{center}
    \includegraphics[width=7cm]{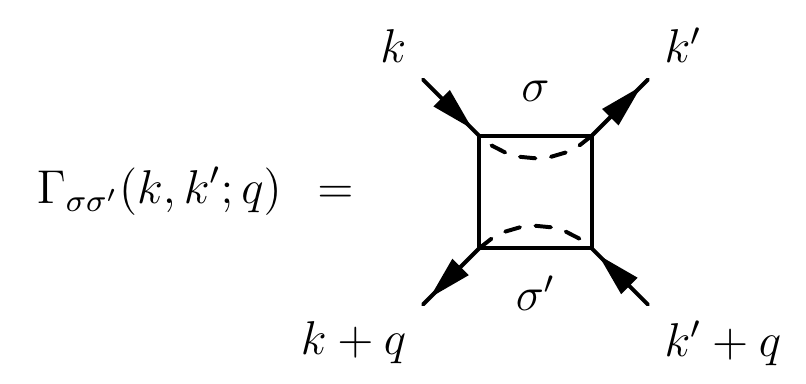}\includegraphics[width=7cm]{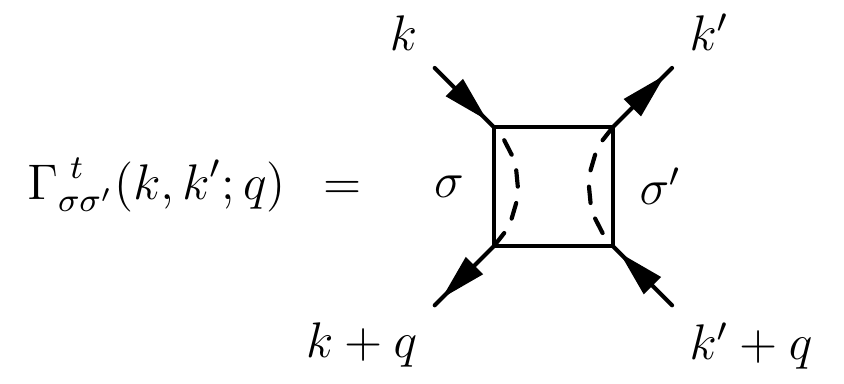}
   \caption{Assignment of the dynamical variables to the two-particle vertex as used in the text. The dashed line expresses the charge and spin conservation of the vertex, which means that the incoming line goes continuously through the vertex and goes out on the other side. The left figure stands for the direct propagation (left to right), the right one for the transposed one (up to down). }  \label{fig:Gamma} 
  \end{center}  
 \end{figure}

Although we formally do not need the two-particle Green function to determine the one-particle one, the perturbation theory for the self-energy contains contributions from  two-particle Green functions, or better two-particle vertices. The \index{two-particle vertex} two-particle vertex, that is, a one-particle irreducible two-particle function,  is related to the two-particle Green function via a defining equation   
\be
G_{\sigma\sigma'}(k,k',q)  = G_{\sigma}(k) G_{\sigma'}(k + q) \left[ \delta(k - k') + \Gamma_{\sigma\sigma'}(k,k';q)G_{\sigma}(k') G_{\sigma'}(k' + q)\right]\,.
\ee
We used a simplified, four-vector, notation of the dynamical variables, momenta and frequencies, the fermionic ones  $k=(\veck,i\omega_{n})$ and the bosonic ones $q=(\vecq,i\nu_{m})$. The two-particle functions obey conservation of momentum and energy and hence they contain only three independent dynamical variables. The attachment of the three variables to four corners of the two-particle vertex is graphically visualized  in Fig.~\ref{fig:Gamma}, where we used the definition of the direct and a transposed two-particle vertices $\Gamma_{\sigma\sigma'}(k,k',q)$ and $\Gamma^{t}_{\sigma\sigma'}(k,k',q)$, respectively.  

The two-particle vertex is only one-particle irreducible. We can introduce a two-particle irreducible vertex  in analogy with the self-energy in order to formally close the dynamical equation for the two-particle vertex. The two-particle analogue of the Dyson equation is a Bethe-Salpeter equation. It is an integral equation generally of the following form  
\be\label{eq:BS-eh}
\Gamma_{\sigma\sigma'} (k,k';q) = \Lambda_{\sigma\sigma'} (k,k';q) - \frac 1{\beta N} \sum_{k''}\Lambda_{\sigma\sigma'} (k,k'';q) G_{\sigma}(k'')G_{\sigma'}(k''+ q) \Gamma_{\sigma\sigma'} (k'',k';q) \,.
\ee

\begin{figure}
  \begin{center}
    \includegraphics[width=11cm]{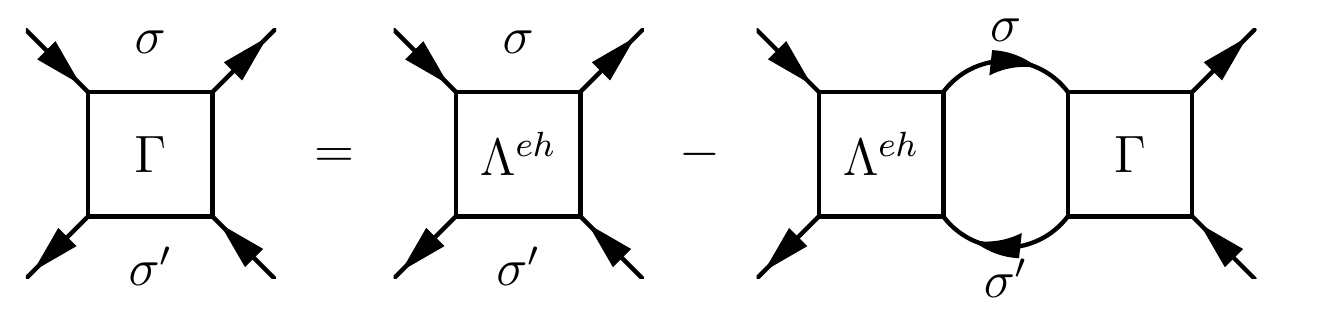}
   \caption{Bethe-Salpeter equation in the horizontal electron-hole channel. The direct and transposed vertices do not mix in this channel.}  \label{fig:BS-eh} 
  \end{center}  
 \end{figure}

Unlike the one-particle irreducibility, the \index{two-particle irreducibility} two-particle irreducibility is not uniquely defined. The irreducibility is best demonstrated graphically with the solid oriented line standing for the one-particle Green function.  The Bethe-Salpeter equation~\eqref{eq:BS-eh} is diagrammatically represented in Fig.~\ref{fig:BS-eh}.  This equation is characterized by a simultaneous propagation of an electron of spin $\sigma$, the upper line, and a hole with spin $\sigma'$, the lower line.  The vertex is electron-hole, two-particle irreducible when cutting simultaneously electron and hole lines does not break the diagram into two disconnected parts. In this case they are horizontal electron and hole lines. Vertex $\Lambda^{eh}$ is the   electron-hole irreducible vertex.  

There are more than one Bethe-Salpeter equations.  Due to the orientation of the lines representing the charge propagation, cutting simultaneously parallel or antiparallel pairs of fermion lines  is not the same operation and leads to a different result with a different conserving dynamical variable. Each of the two-particle irreducibility can actually be characterized by a bosonic dynamical variable that is conserved, that is, it is the same whenever we perform the cut. It is $q=(\vecq,i\nu_{m})$ in Eq.~\eqref{eq:BS-eh}  for vertex $\Gamma_{\sigma\sigma'} (k,k';q)$  with the attachment of the variables defined in Fig.~\ref{fig:Gamma}. 
 
\begin{figure}
  \begin{center}
    \includegraphics[width=15cm]{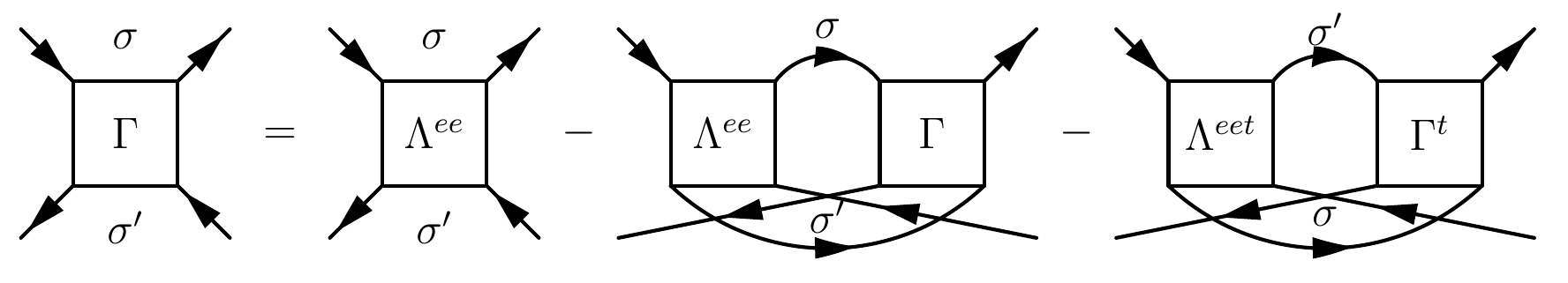}
   \caption{Graphical representation of the Bethe-Salpeter equation with the simultaneous propagation of two electrons. The direct and transposed vertices contribute with the same diagram with only interchanged spins in the intermediate propagation.  }  \label{fig:BS-ee} 
  \end{center}  
 \end{figure}

Another option of the horizontal two-particle propagation is in connecting the vertices by parallel lines. The corresponding \index{Bethe-Salpeter equation} Bethe-Salpeter equation is graphically represented in Fig.~\ref{fig:BS-ee}. Unlike Eq.~\eqref{eq:BS-eh}, the direct and transposed vertices are now mixed up. We can see that the transposed vertices are connected in the same way as the direct ones, only the upper and lower interconnecting propagators are interchanged. Checking the conservation laws in the vertices we easily find that the conserving dynamical variable in this channel is $Q=q + k'+ k$.  Vertex $\Lambda^{ee}$ is electron-electron irreducible and differs from vertex $\Lambda^{eh}$, being irreducible in the electron-hole channel of Fig.~\ref{fig:BS-eh}.  

\begin{figure}
  \begin{center}
    \includegraphics[width=15cm]{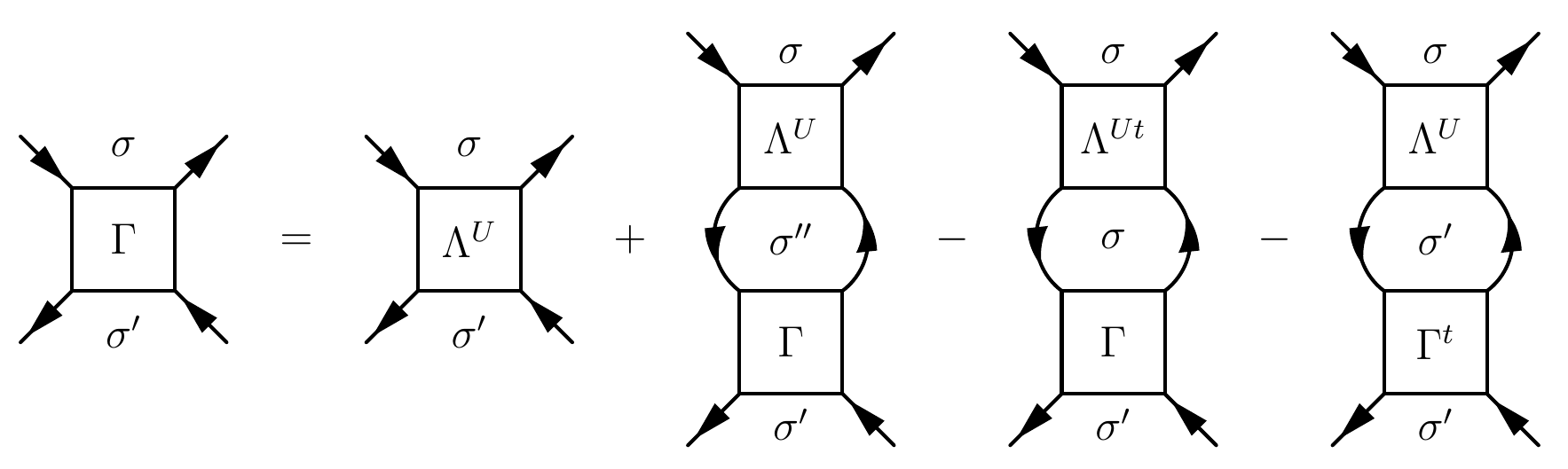}
   \caption{Bethe-Salpeter equation for the vertical (transposed) propagation of the electron and the hole with the same spin. The double primed variables are dummy integration variables. Notice, that vertex $\Gamma_{\uparrow\downarrow}$ is coupled to $\Gamma_{\uparrow\uparrow}$ and $\Gamma_{\downarrow\downarrow}$.}  \label{fig:BS-U} 
  \end{center}  
 \end{figure}

There is yet another \index{Bethe-Salpeter equation} Bethe-Salpeter equation and another two-particle irreducibility. It is an electron-hole irreducibility  propagating vertically (in the transposed sense) an electron and a hole carrying the same spin. It differs from the horizontal electron-hole propagation in Fig.~\ref{fig:BS-eh} where the spins of the particle and the hole are independent. The graphical representation of such a Bethe-Salpeter equation is shown in Fig.~\ref{fig:BS-U}. The double primed variable indicates that it is summed over. The direct and transposed vertices are mixed in this channel in a more complex way than in Fig.~\ref{fig:BS-ee}. The sign at each diagram in the graphical representation accounts for the anticommuting character of the one-particle operators. The conserving dynamical variable in this vertical channel is $\Delta k = k - k'$.  The irreducible vertex $\Lambda^{U}$ differs from both preceding vertices $\Lambda^{eh}$ and $\Lambda^{ee}$. 

The full two-particle vertex $\Gamma$ is the same in all three Bethe-Salpeter equations. We thus have three inequivalent representations of the two-particle vertex via irreducible ones.  Simultaneously we could construct three Bethe-Salpeter equations for the transposed vertex $\Gamma^{t}$. We obtain for the equal-spin vertex  $\Gamma^{tot}_{\sigma\sigma}(k,k';q) = \Gamma_{\sigma\sigma}(k,k';q) + \Gamma^{t}_{\sigma\sigma}(k,k';q)$ with the symmetry relation $\Gamma^{t}_{\sigma\sigma}(k,k';q) = - \Gamma_{\sigma\sigma}(k,k + q;k'-k)$.   

\section{Baym-Kadanoff construction of the renormalized perturbation theory} 

Perturbation theory can be constructed independently for Green functions of any order. The one-particle Green function is an input to the perturbation theory for higher-order Green functions. It is only expected to obey the desired analytic properties. To keep the whole perturbation theory for all types of Green functions consistent and conserving we must relate the one-particle Green function with the higher-order ones.  The renormalizations of the one-particle and higher-order Green functions cannot be done independently. The canonical way to construct renormalized and conserving approximations was proposed by Baym and Kadanoff \cite{Baym:1961aa,Baym:1962aa}.

\subsection{Generating Luttinger-Ward functional and irreducible functions}

The basic element od the perturbation theory is  the one-particle Green function. Its behavior is essential since it enters all physical quantities. It is hence the primary task of the perturbation theory to evaluate the one-particle Green function. It is sufficient to know the self-energy since the Dyson equation determines the Green function from the self-energy. The fundamental quantity of the perturbation theory becomes then the self-energy. The principal assumption of the Baym-Kadanoff approach is that the self-energy $\Sigma$ can be expressed as a functional of the renormalized one-particle Green function $G$ and the bare interaction $U$, $\Sigma[G,U]$.   

The conserving character of the perturbation theory demands the existence of a generating functional from which we can determine all physical quantities via functional derivatives. The Baym-Kadanoff self-energy is obtained from the  \index{Luttinger-Ward functional} Luttinger-Ward generating functional $\Phi[G,U]$. The thermodynamic functional of the Baym-Kadanoff theory is the grand potential in the Matsubara formalism  
\bdm\label{eq:Omega-Phi}
\frac 1N\Omega[G,\Sigma] = -\frac 1{\beta N}\sum_{\sigma}\sum_{\omega_{n},\mathbf{k}} e^{i\omega_{n}0^{+}}\left\{\ln\left[i\omega_{n} + \mu_{\sigma} - \epsilon(\mathbf{k}) - \Sigma_{\sigma}(\mathbf{k},i\omega_{n}) \right] + G_{\sigma}(\mathbf{k},i\omega_{n}) \Sigma_{\sigma}(\mathbf{k},i\omega_{n}) \right\}  + \Phi[G,U] \,.
\edm 
The equilibrium value of functional $\Omega[G,\Sigma]$ is stationary with respect to the variations of its functional variables, $\Sigma_{\sigma}(\mathbf{k},i\omega_{n}), G_{\sigma}(\mathbf{k},i\omega_{n})$. Functional derivatives of the Luttinger-Ward functional with respect to Green function $G_{\sigma}(\mathbf{k},i\omega_{n})$ lead to the irreducible functions. The first and second derivatives lead to the self-energy and the two-particle irreducible vertex 
\begin{align}\label{eq:Phi-derivatives}
\Sigma_\sigma(k) = \frac{\delta\Phi[G,U]}{\delta G_{\sigma}(k)}\ , &\quad \Lambda_{\sigma\sigma'}(k,k';q) = \frac{\delta \Sigma_{\sigma}(k,k')}{ \delta G_{\sigma'}(k'+ q,k + q)} \,.
\end{align}
Stationarity of the grand potential $\Omega[G,\Sigma]$ with respect to the variations of $\Sigma_\sigma(\veck,i\omega_{n})$ leads to the Dyson equation for the one-particle Green function $G_{\sigma}(\mathbf{k},i\omega_{n})$ renormalized by the self-energy $\Sigma_{\sigma}(\mathbf{k},i\omega_{n})$ from Eq.~\eqref{eq:Phi-derivatives}. Notice that combinations of the independent variables in the second derivative of Eq.~\eqref{eq:Phi-derivatives} lead to different two-particle irreducible vertices.  It holds for all  higher-order irreducible functions that are not  unique and depend on the way we use the internal variables out of equilibrium in the functional derivatives. The  Baym-Kadanoff perturbation theory is hence fully one-particle renormalized. The higher-order Green and irreducible functions do not enter the generating functional and are not determined self-consistently.  

The fundamental quantity of the Baym-Kadanoff construction is the Luttinger-Ward functional. Its existence is guaranteed by interchangeable second derivatives in \index{ $\Phi$-derivable theories} $\Phi$-derivable theories. The Luttinger-Ward functional can be formally obtained from the bare perturbation theory via a Legendre transformation by replacing the inverse of the bare propagator by the fully renormalized Green function
\be
\Phi[G,U] =  -k_{B}T\ln\mathcal{Z}[G^{(0)-1},U] - \int d\ \bar{1}\left( G^{(0)-1}(1,\bar{1}) -   G^{-1}(1,\bar{1})\right) G(\bar{1},1^\prime) \,.
\ee
It is, however, only a formal representation that does not help us find the functional dependence of the Luttinger-Ward functional on the full Green function. The Luttinger-Ward functional is not the primary object of interest in the Baym-Kadanoff construction. It is replaced by the self-energy the contributions to which from the perturbation expansion are better controlled.

\subsection{Schwinger-Dyson equation, Ward identity \& Schwinger field theory}

\begin{figure}
  \begin{center}
    \includegraphics[width=14cm]{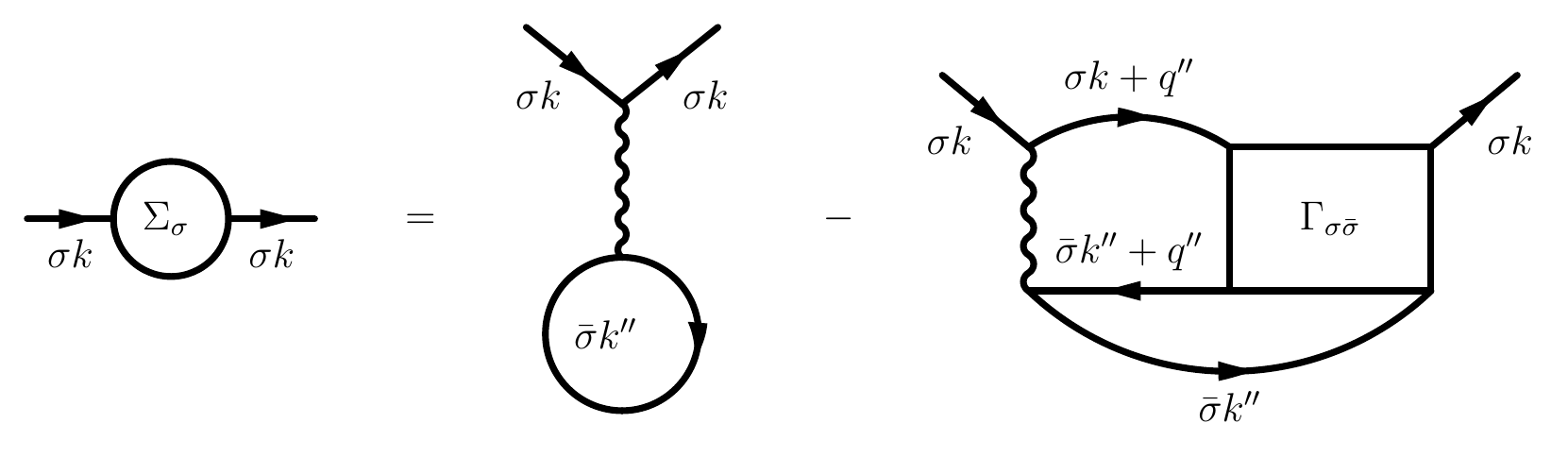}
   \caption{ Graphical representation of the \index{Schwinger-Dyson equation} Schwinger-Dyson equation of the Hubbard model. The first term on the right-hand side is the Hartree term and the second is the contribution from the two-particle vertex $\Gamma$. We kept the variables of one-particle propagators. The double-prime indices are summed (integrated) over. The bar denotes the inverse value.}  \label{fig:SDE} 
  \end{center}  
 \end{figure}

The contributions from the perturbation theory to the self-energy functional can be regrouped so that one singles out the static Hartree contribution and represent the dynamical part via the two-particle vertex. Its graphical representation is plotted in Fig.~\ref{fig:SDE}. Mathematically it is represented for the Hubbard-like models as       
\bdm \label{eq:SDE-1}
\Sigma_\sigma[G,U](k) = \frac{U}{\beta N}\sum_{k'} G_{\bar{\sigma}}(k')\left[1  -\frac{1}
  {\beta N}\sum_{q}\Gamma_{\sigma\bar{\sigma}}[G,U](k,k';q)
  G_\sigma(k+q)G_{\bar{\sigma}}(k'+q)\right] \,.
\edm
The two-particle vertex $\Gamma[G,U]$ contains the unspecified dependence on the one-particle Green function that has to be also determined from the perturbation theory. Its form is, however, fully determined by the Luttinger-Ward functional $\Phi[G,U]$, if known. 

The two-particle vertex can be represented by the two-particle irreducible vertices in the corresponding Bethe-Salpeter equations. The two-particle irreducible functions do not enter the Luttinger-Ward functional, but they are determined from its functional derivatives. The second functional derivative in Eq.~\eqref{eq:Phi-derivatives} is a differential form of the \index{Ward identity} Ward identity  that is a sufficient microscopic condition that guarantees validity of the continuity equation of the macroscopic system. The differential Ward identity  can be integrated to \cite{Janis:2003ab,Krien:2017aa}   
\be\label{eq:WI-Integral}
\Sigma_{\sigma}(k + q) - \Sigma_{\sigma}(k)   = \frac 1{\beta N}\sum_{k'}\Lambda_{\sigma\sigma}^{eh}(k,k';q) \left[ G_{\sigma}(k' + q) -  G_{\sigma}(k')\right] \
\ee
\begin{figure}
  \begin{center}
    \includegraphics[width=7cm]{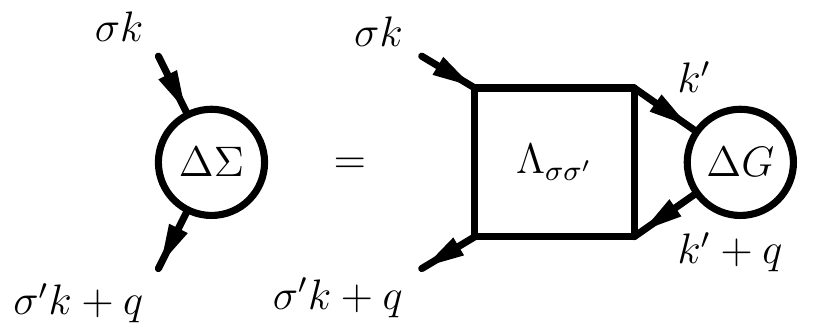}
   \caption{\index{Ward identity} Ward identity matching a difference $\Delta\Sigma$ of the self-energy with the two-particle irreducible vertex $\Lambda$.}  \label{fig:WI} 
  \end{center}  
 \end{figure}
and is plotted graphically in Fig.~\ref{fig:WI}. If we use one of the Bethe-Salpeter equations to determine the full two-particle vertex and the differential form of the Ward identity for the two-particle irreducible vertex in the Schwinger-Dyson equation we end up in an integro-differential functional equation for the self-energy
\be \label{eq:Schwinger-FT}
\Sigma = U \left\langle G - G\left[1 + 	\frac{\delta\Sigma}{\delta G}GG\star\right]^{-1}\frac{\delta\Sigma}{\delta G}GG\right\rangle \,,
\ee
where the star denotes the appropriate two-particle convolution from the Bethe-Salpeter equation and the angular brackets stand for the summation over the one-particle integration variables. Equation~\eqref{eq:Schwinger-FT} represents the \index{Schwinger field theory} Schwinger field theory for the self-energy \cite{Schwinger:1949aa}. Unlike the Luttinger-Ward functional, the Schwinger field theory explicitly utilizes the exact properties of both the one and two-particle Green functions. The solution of Eq.~\eqref{eq:Schwinger-FT} can be reached only iteratively via an expansion in the interaction strength, which is equivalent to the standard perturbation theory for the thermodynamic potential. 

Although the Luttinger-Ward functional guarantees a conserving theory, none of the self-energy functionals $\Sigma[G,U]$ determined from the perturbation theory is fully conserving. A fully conserving theory should also obey conservation of the particle interaction. Namely, the strength of the particle interaction is fully generated by the present electrons in the system.  This conservation is mathematically expressed as a sum rule  \cite{Janis:1998aa} 
\be
\frac{\partial\Omega(U,\mu_{\mathbf{i}\sigma}) }{\partial U} = \sum_{\mathbf{i}}\left[\frac{\delta^{2}\Omega}{\delta \mu_{i\uparrow}\delta \mu_{i\downarrow}}  + \frac{\delta\Omega}{\delta \mu_{i\uparrow}}  \frac{\delta\Omega}{\delta \mu_{i\downarrow}}\right] = \sum_{i}\left\{ \frac{k_{B}T}4 \left[\kappa_{ii} - \chi_{ii}\right] + n_{i\uparrow}n_{i\downarrow}\right\} \,,
\ee
where $\mu_{i\sigma}$ are the local spin-dependent chemical potentials and $\kappa_{ii}$ and $\chi_{ii}$ are local compressibility and susceptibility, respectively. This identity expresses that the charges behind the Coulomb repulsion  are entirely carried by the present electrons. The meaning of this identity is that mass and charge of the electrons  cannot be separated.   

The above sum rule can be extended to a gauge transformation by making interaction and mass of the electrons dynamical. The corresponding  generalized Ward identity  \index{Ward identity-generalized} then reads
\be\label{eq:WI-charge}
\underbrace{
\frac{\delta \Phi[U,G]}{\delta U(\mathbf{q},i\nu_{m})}}_{\text{Schwinger-Dyson}} = \underbrace{-\frac{1}{ \beta N}\sum_{\mathbf{k},\omega_{n}}\frac{\delta G_{\sigma}(\mathbf{k} + \mathbf{q},i\omega_{n} + i\nu_{m})}{\delta \mu_{-\sigma}(\mathbf{k},i\omega_{n})} \,.
}_{\text{Ward}}
\ee
Both sides of this dynamical equation should lead to the same  two-particle correlation function. The left-hand side results in a correlation function obtained from the vertex of the Schwinger-Dyson equation, while the right-hand side determines a correlation function from the Bethe-Salpeter equation with the irreducible vertex satisfying the differential Ward identity of Eq.~\eqref{eq:Phi-derivatives}. Ward identity~\eqref{eq:WI-charge} complies with the Schwinger-Dyson equation only if the irreducible vertex obeys an integro-differential  functional equation \cite{Janis:2017aa}
\bdm
\Lambda^{eh}_{\sigma\bar{\sigma}} =  U - U\left[ 1 + G_{\sigma}G_{\bar{\sigma}}\Lambda^{eh}_{\sigma\bar{\sigma}}\star\right]^{-1}G_{\sigma}\left\{\Lambda^{eh}_{\sigma\bar{\sigma}}%\phantom{\frac 12}
 + G_{\bar{\sigma}}\frac{\delta\Lambda^{eh}_{\sigma\bar{\sigma}}}{\delta G_{\bar{\sigma}}}\right\}
\left[ 1 + \star G_{\sigma}G_{\bar{\sigma}}\Lambda^{eh}_{\sigma\bar{\sigma}}\right]^{-1}\circ G_{\bar{\sigma}} \,, 
\edm
where we denoted $\bar{\sigma} = -\sigma$. Finding its solution is equivalent to solving the Schwinger field theory. Consequently, the Schwinger-Dyson equation~\eqref{eq:SDE-1}  and the differential Ward identity from Eq.~\eqref{eq:Phi-derivatives} can never be fully satisfied in approximate solutions in the many-body perturbation theory.

\subsection{Simple approximations: Hartree, RPA, FLEX}

Having set the general rules for the renormalized pertrurbation theory we proceed to use them to derive approximate solutions. The simplest is the \index{Hartree approximation} Hartree approximation. It is a static  mean-field theory that can be derived from the grand potential of Eq.~\eqref{eq:Omega-Phi} by choosing the simplest  Luttinger-Ward functional 
$$\Phi_{Hartree}[G,U]= \frac U{ N^{2}}\sum_{\veck,\veck'}\frac 1{\beta^{2}}\sum_{\omega_{n}\omega_{n'}}e^{i(\omega_{n} + \omega_{n'})0^{+}} G_{\uparrow}(\veck,i\omega_{n})G_{\downarrow}(\veck',i\omega_{n'})\,.
$$
The resulting Hartree self-energy is 
\be\label{eq:SE-Hartree}
\Sigma_{\sigma}[G,U] = \frac U{\beta N} \sum_{\veck,\omega_{n}} e^{i\omega_{n}0^{+}} G_{\bar{\sigma}}(\veck,i\omega_{n})  = Un_{\bar{\sigma}}\,,
\ee
where $n_{\sigma}$ is the spin density. Consequently, $\Gamma_{\sigma\sigma'}[G,U]= 0$ in the Schwinger-Dyson equation~\eqref{eq:SDE-1}. The resulting theory is not trivial because the thermodynamic properties are derived from the Ward identity and the irreducible two-particle vertex from Eq.~\eqref{eq:Phi-derivatives}, $\Lambda_{\uparrow\downarrow}[G,U]= U$.  This irreducible vertex leads to a nontrivial full two-particle vertex 
\be\label{eq:Gamma-Hartree}
\Gamma_{\uparrow\downarrow} (\vecq,i\nu_{m}) = \frac U{1 + U\phi_{\uparrow\downarrow}(\vecq,i\nu_{m})}
\ee
when we use it in the Bethe-Salpeter equation~\eqref{eq:BS-eh}. We introduced a two-particle (electron-hole)  bubble
\bdm\label{eq:phi-eh}
\phi_{\uparrow\downarrow}(\vecq,i\nu_{m}) = \frac 1{\beta N}\sum_{\veck,i\omega_{n}} G_{\uparrow}(\veck + \vecq,i\omega_{n} + i\nu_{m}) G_{\downarrow}(\veck,i\omega_{n})  = - \frac 1N\sum_{\veck}\int_{-\infty}^{\infty} \frac{dx}{\pi} f(x)  \left[G(\veck + \vecq, x + \omega_{+}) + G(\veck - \vecq,x -  \omega_{+}) \right] \Im G(\veck, x_{+})\,. 
\edm
We denoted $\omega_{\pm} = \omega \pm i0^{+}$. Vertex $\Gamma_{\uparrow\downarrow}$ from Eq.~\eqref{eq:Gamma-Hartree} generates the thermodynamic behavior of the Hartree mean-field approximation.  We can further use it in the Schwinger-Dyson equation to derive a new dynamical, spin-symmetric self-energy  
\bdm\label{eq:SE-RPA}
 \Sigma^{Sp}(\veck,\omega_{+}) =  \frac{U}{N}\sum_{\vecq}P\int_{-\infty}^{\infty} \frac{dx}{\pi}\left\{b(x) G(\veck + \vecq, \omega_{+} + x) \Im\left[\frac{1}{1 + U \phi(\vecq,x_{+})} \right]     
% \right. \\ \left.
 - \frac{f(x + \omega)}{1 + U\phi(\vecq, x_{-}) }\Im{G}(\veck + \vecq,x + \omega_{+})\right\} \,,
\edm
where the Green functions are renormalized by the Hartree self-energy from Eq.~\eqref{eq:SE-Hartree}.  Only the linear term reproduces the Hartree self-energy. The self-energy $\Sigma^{Sp}(\veck,\omega_{+})$ from this equation represents the \index{random-phase approximation} random-phase approximation (RPA). The RPA is not $\Phi$-derivable, since the Green functions in the Schwinger-Dyson equation are not one-particle self-consistent. If we replace the Hartree self-energy with $\Sigma^{Sp}(\veck,\omega_{+})$ from Eq.~\eqref{eq:SE-RPA} and replace the Hartree propagators in Eqs.~\eqref{eq:phi-eh} and on the right-hand side of Eq.~\eqref{eq:SE-RPA} by the fully renormalized one
\be
G(\veck, \omega_{+}) = \frac 1{\omega_{+} + \mu -\epsilon(\veck) - \Sigma^{Sp}(\veck,\omega_{+})}\,,
\ee
the self-energy $\Sigma^{Sp}(\veck,\omega_{+})$ then generates a \index{fluctuation-exchange} fluctuation-exchange approximation (FLEX). This approximation is fully one-particle self-consistent and its Luttinger-Ward functional reads \cite{Janis:1998aa}
\be
\Phi_{FLEX}[G,U]=\frac 1N\sum_{\vecq}P\int_{-\infty}^{\infty}\frac{d\omega}{\pi} b(\omega)\Im\left[U\phi_{\uparrow\downarrow}(\vecq,\omega_{+}) -  \ln\left(1 + U \phi_{\uparrow\downarrow}(\vecq,\omega_{+})\right)\right]\,.
\ee 

These simple approximations derived within the Baym-Kadanoff construction and the generating Luttinger-Ward functional demonstrate the inability to obey simultaneously the Ward identity and the Schwinger-Dyson equation. They also demonstrate that the Ward identity is important for deriving thermodynamic properties averaged over Matsubara frequencies, while the Schwinger-Dyson equation is responsible for the dynamics and dynamical effects. The level of self-consistency in the Schwinger-Dyson equation may significantly determine the quality of the approximations. The RPA shares the thermodynamic properties with the Hartree approximation and introduces the corresponding dynamical  corrections in the spectral function. The FLEX theory changes both the thermodynamic and also dynamical properties of the Hartree approximation but the two are derived from different vertex functions and hence, inconsistent. An alternative approach should be chosen to reconcile the thermodynamic and spectral properties of self-consistent theories.

\section{Two-particle approach \& two-particle renormalizations}

The Baym-Kadanoff approach leads directly to the Schwinger-Dyson equation determining the one-particle self-energy. It contains only one-particle functions and includes only \index{one-particle renormalizations} one-particle renormalizations. The two-particle functions are derived indirectly. There are two vertex functions to a single self-energy. One is the vertex extracted from the Schwinger-Dyson equation and the other is constructed from a Bethe-Salpeter equation with the two-particle irreducible vertex related to the self-energy via the Ward identity. The two vertices are different in all approximate schemes and the critical behavior cannot be determined uniquely in such a construction.     
 
The critical behavior connected with a phase transition is real and unique. That is why it must also be identified uniquely in theoretical models. It is then of utmost importance to have only a single two-particle vertex that would lead to the experimentally observed critical behavior. Moreover, a direct \index{two-particle self-consistency} two-particle self-consistency is needed to suppress unphysical and spurious critical behavior of weak-coupling approximations and to gain a full control of the critical behavior.  Since we cannot satisfy simultaneously the Ward identity and the Schwinger-Dyson equation we should accept the existence of two self-energies to a single two-particle vertex. The generating functional of such an approach cannot be a thermodynamic potential or the self-energy but rather the two-particle irreducible vertex from the singular Bethe-Salpeter equation of the studied critical behavior. It will be determined from the perturbation theory. One self-energy derived from this vertex will represent a \index{thermodynamic order parameter} thermodynamic order parameter and will obey the Ward identity. The other self-energy will be derived from the Schwinger-Dyson equation without changing the critical behavior of the two-particle vertex and the order parameter.

\subsection{Symmetry-breaking field - odd and even functions}

The basic characteristic quantity that distinguishes the two self-energies derived from a single two-particle vertex is the \index{symmetry-breaking field} symmetry-breaking field in which the linear-response theory breaks down at the critical point.  It is related to the order parameter in the low-temperature ordered phase via a Legendre transformation.  We separate quantities with even and odd symmetry with respect to the reversal of this field. The order parameter is a typical quantity with odd symmetry. Let us assume, e.~g., a symmetry-breaking field $\eta^{\perp}$ generating spin-flip processes. Its Legendre conjugate order parameter is 
\be
\Delta_{\uparrow\downarrow} = \frac 1N\sum_{i}\left\langle c_{\downarrow}^{\dagger}({\bf R}_{i})c_{\uparrow}^{\phantom{\dagger}}({\bf R}_{i})\right\rangle \,.
\ee
It describes the  transverse magnetic order. Both the field and the order parameter are generally complex. It is sufficient to keep only the leading terms in the symmetry-breaking field to describe qualitatively correctly the critical behavior. Only the symmetric two-particle functions remain then relevant in the critical region.  We hence define a  symmetric two-particle propagator with a general external field $\eta$  
\begin{multline}
\left[G(\veck,i\omega_{n}) G(\veck + \vecq,i\omega_{n} + i\nu_{m})\right]_{\eta} =  \frac 12\left[G_{\uparrow}(\veck,i\omega_{n};\eta)G_{\downarrow}(\veck + \vecq,i\omega_{n} + i\nu_{m};\eta)
\right. \\ \left.
 +\ G_{\downarrow}(\veck,i\omega_{n};-\eta) G_{\uparrow}(\veck + \vecq,i\omega_{n} + i\nu_{m};-\eta) \right] \,,
\end{multline}
where index $\eta$ denotes the field with respect to which the symmetrization is performed. 

The one-particle functions must be considered both with odd and even symmetries. The  odd and even components  of the one-particle Green function are   
\begin{subequations}\label{eq:G-even-odd}
\begin{align}
\Delta_{\eta} G(\veck,i\omega_{n}) &= \frac 12\left[G_{\sigma}(\veck,i\omega_{n};\eta) - G_{\bar{\sigma}}(\veck,i\omega_{n};-\eta) \right] \,,
 \\
\bar{G}_{\eta}(\veck,i\omega_{n}) &= \frac 12\left[G_{\sigma}(\veck,i\omega_{n};\eta) + G_{\bar{\sigma}}(\veck,i\omega_{n};-\eta) \right] \,. 
\end{align}
\end{subequations}
The \index{odd propagator} odd propagator will determine the thermodynamic order parameter while the \index{even propagator} even propagator will enter the Schwinger-Dyson equation to complete the description of the dynamical and spectral behavior of the chosen approximation without affecting its thermodynamic critical behavior.

\subsection{Two-particle self-consistency and charge renormalization}

The main qualitative difference between the two-particle approach and the Baym-Kadanoff construction is a two-particle self-consistency introduced via a renormalization of the two-particle irreducible vertices. We know that there are three two-particle irreducible vertices generating three independent Bethe-Salpeter equations for a single full two-particle vertex. A two-particle self-consistency in the perturbation theory for the two-particle irreducible vertices can be expressed functionally $\Lambda^{\alpha}_{ren}\left[G,\Gamma \right]$. It means that we replaced the bare interaction $U$ by the full two-particle vertex $\Gamma$. The approximate functional dependence must be, however, chosen carefully to avoid multiple summations of diagrams.  The \index{two-particle self-consistency} two-particle self-consistency is introduced by an equation eliminating the full vertex in favor of the irreducible ones and the bare interaction
\be
F\left(\left\{ \Lambda^{\alpha}_{ren}\left[G,\Gamma \right]\right\}_{\alpha=1}^{l},U\right) = 0 \,.
\ee
Replacing the bare interaction of the Baym-Kadanoff perturbation theory with the two-particle vertex $\Gamma$ in the two-particle approach, that is a replacement $\Lambda^{\alpha}[G,U]\to\Lambda^{\alpha}_{ren}\left[G,\Gamma \right]$, corresponds to  \index{charge renormalization} charge renormalization since the bare interaction is proportional to the square of the charge of the electrons.

Straightforward two-particle renormalizations are induced by the parquet approach and its simplifying variants introduced in the many-body perturbation theory by De Dominicis and Martin \cite{DeDominicis:1964aa,DeDominicis:1964ab}. Review of the recent developments of the parquet theory can be found in Refs.~\cite{Bickers:1991ab,Rohringer:2018aa}.  Its main idea is to use the ambiguity in the definition of the two-particle irreducibility. The reducible vertex in one scattering channel becomes irreducible in the others. If $\Lambda^{\alpha}$ is the irreducible vertex and $\mathcal{K}^{\alpha}$ its  corresponding reducible counterpart in scattering channel $\alpha$ then the parquet equations can be written as 
\be\label{eq:parquet0}
\Gamma = \Lambda^{\alpha} + \mathcal{K}^{\alpha} = \mathcal{I}_{l} + \sum_{\alpha=1}^{l}\mathcal{K}^{\alpha} \,,
\ee
where $\mathcal{I}_{l}$ is the \index{fully irreducible vertex} fully irreducible vertex in the selected $l$ scattering channels in the parquet theory. The full vertex $\Gamma$ can be excluded from Eq.~\eqref{eq:parquet0} and the parquet equations are closed for either the irreducible $\Lambda^{\alpha}$ or the reducible $\mathcal{K}^{\alpha}$ vertices.  The input to the parquet equations are the one-particle propagators $G_{\sigma}$ and the fully irreducible vertex $\mathcal{I}_{l}$. The latter is usually replaced by the bare interaction, $\mathcal{I}_{l}=U$. It is important to mention that the parquet decomposition does not hold always, but only when the overlap of the sets of the reducible diagrams from different channels is empty \cite{Janis:2009aa}.

\subsection{Order parameter and mass renormalization}

The perturbation theory for the two-particle vertex functions is initially independent of the perturbation theory for the one-particle Green functions.  The one-particle Green functions enter the perturbation theory for the two-particle functions only as an input with the expected analytic properties. One needs, however, to match the perturbation theories for one and two-particle Green function in order to keep the approximations conserving. We already know that we cannot simultaneously  obey all the exact relations between the one and two-particle functions. We must hence decide which relation in which situation is relevant and must not be neglected. Approximate fulfillment of the exact relations need not have a qualitative impact on the perturbative solutions. The critical regions of the phase transitions are, however, different. We must be very careful in breaking the relations between the one and two-particle functions there. 

\begin{figure}
  \begin{center}
    \includegraphics[width=6.5cm]{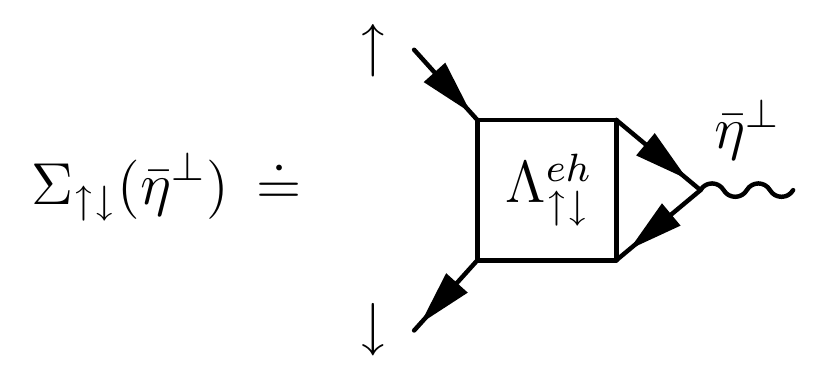}
   \caption{The leading linear order of the contribution from the symmetry-breaking field, $\eta^{\perp}$ in this case, to the self-energy. }  \label{fig:SBF} 
  \end{center}  
 \end{figure}

The first step was already made by separating the one-particle functions with odd and even symmetry with respect to the \index{symmetry-breaking field} symmetry-breaking field of the studied critical behavior. Thermodynamic consistency demands that the one-particle order parameter starts to grow from zero at the critical point of the two-particle response function. This is achieved only when the Ward identity is obeyed.  Qualitative consistency is reached already when the Ward identity is satisfied in the leading linear order in the symmetry-breaking field.  The linear dependence of the self-energy on perturbation $\eta^{\perp}$ is plotted in Fig.~\ref{fig:SBF}.  This is the \index{odd self-energy} odd self-energy and it will be related to the two-particle vertex via the \index{Ward identity} Ward identity. 

The Ward identity linearized with respect to the static homogeneous field $\eta^{\perp}$ as in Fig.~\ref{fig:SBF} is 
\be
\Delta_{\eta}\Sigma(\veck,i\omega_{n}) = \frac 1{\beta N}\sum_{\veck',\omega_{n'}}\Lambda^{eh}_{\eta}(\veck,i\omega_{n},\veck',i\omega_{n'};0,0)\Delta_{\eta} G(\veck',i\omega_{n'}) \,,
\ee
where we used the definitions of Eqs.~\eqref{eq:G-even-odd}. 

The Ward identity is a microscopic relation guaranteeing macroscopic conservation laws. It has no direct relation to the microscopic quantum dynamics. The latter is governed by the Schwinger-Dyson equation. The dynamics is not directly connected with the thermodynamic critical behavior. The only consistency requirement is that the vertex in the Schwinger-Dyson equation is built from the same irreducible vertex used in the Ward identity via the corresponding Bethe-Salpeter equation. We hence demand that the \index{even self-energy} even self-energy be determined from the following \index{Schwinger-Dyson equation} Schwinger-Dyson dynamical equation
\bdm%\begin{multline}
 \bar{\Sigma}_{\eta}(\veck,i\omega_{n}) =  \frac U2 n -  \frac{U}{N^{2}}\sum_{\veck'\vecq}\frac 1{\beta^{2}}\sum_{\omega_{n'}\nu_{m}}\bar{G}_{\eta}(\veck',i\omega_{n'}) \bar{G}_{\eta}(\veck' + \vecq,i\omega_{n'} + i\nu_{m}) \\
 \times\Gamma_{\eta}(\veck,i\omega_{n},\veck',i\omega_{n'};\vecq,i\nu_{m})\bar{G}_{\eta}(\veck + \vecq,i\omega_{n} + i\nu_{m}) \,,
\edm%\end{multline}
where $n$ is the total charge density. 

The self-energy is split into odd and even parts. The odd part plays the role of the order parameter and is determined from the \index{Ward identity-linearized} linearized Ward identity. The even part obeys the \index{Schwinger-Dyson equation} symmetrized Schwinger-Dyson equation and is responsible for the dynamical behavior and the spectral properties. This general construction of two self-energies is an extension of the Hartree thermodynamics and the RPA dynamics.  At the end, however, all the physical quantities must be determined from the full one-particle Green function with the total self-energy $\Sigma(\veck,i\omega_{n}) = \Delta\Sigma(\veck,i\omega_{n}) + \bar{\Sigma}(\veck,i\omega_{n})$. The way we separate the odd and even self-energy depends on the critical behavior and the controlling symmetry-breaking field. The construction is fitted just to the investigated critical behavior. The order parameter can be anomalous with an anomalous Green function in the ordered phase breaking macroscopic conservation laws. 
%
%\be
%G_{\sigma}(\veck,i\omega_{n}) = \frac 1{i\omega_{n} + \mu - \sigma\Delta\Sigma(\veck,i\omega_{n}) -  \bar{\Sigma}(\veck,i\omega_{n})} \,.
%\ee

\section{Mean-field theory with a two-particle self-consistency}

The main difference between the Baym-Kadanoff construction and the two-particle approach is the selection of the functional generating the perturbation theory. The former starts with the Luttinger-Ward functional and takes the self-energy as the fundamental object of the perturbation theory.  The latter makes the two-particle irreducible vertices the central objects of the perturbation theory. The two approaches formally coincide in the high-temperature phase with no odd self-energy. They differ in the critical regions of the phase transitions where the Ward identity must be obeyed, at least in the leading order of the symmetry-breaking field. This is guaranteed only in the direct perturbation theory for the two-particle functions.

The perturbation theory for the self-energy is much easier to handle than the theory of the two-particle irreducible vertices. It is evident from the equations in which they enter the one and the two-particle Green functions. The one-particle Green function is determined from an algebraic Dyson equation, while the two-particle vertex from integral Bethe-Salpeter equations.  It means that it is still a cumbersome way to determine the physical response functions from the generating two-particle irreducible functions. The major problem of the two-particle functions is their complexity with three independent dynamical variables and an unknown analytic structure. This is the main hurdle in the application of the two-particle approach. That is why most two-particle theories are based on heavy numerics \cite{Tam:2013aa,Li:2019aa,Eckhardt:2020aa}. 
     
The starting point of the perturbation theory is a reliable mean-field approximation with static renormalizations of the input parameters as discussed in Sec.~\ref{sec:StaticRen}. The  Baym-Kadanoff approach  leads only to a weak-coupling mean-field approximation. Moreover, due to the lack of the  two-particle self-consistency, it contains spurious transitions. One expects that this flaw will be removed by the two-particle approach. The first task of the two-particle perturbation theory is then to construct a mean-field theory with a two-particle self-consistency.

\subsection{Reduced parquet equations}

We discussed that the most straightforward way to introduce a two-particle self-consistency is to use the parquet equations interconnecting the two-particle irreducible/reducible vertices. The two-particle approach must be able to reach the critical region of the weak-coupling theory in which the two-particle self-consistency suppresses the spurious crossing of the critical point. Unfortunately, the full set of the parquet equations misses the critical point of the RPA and is unable to reach the Kondo regime in the SIAM \cite{Janis:2006ab}. That is why we developed a simplified analytic theory with reduced parquet equations that we used to construct a mean-field theory with a two-particle self-consistency \index{two-particle self-consistency} \cite{Janis:2007aa,Janis:2017aa,Janis:2019aa}.   

The first step in the reduction of the parquet equations is to select the Bethe-Salpeter equation with a critical point of the unrenormalized perturbation theory. It is the equation with \index{multiple electron-hole scatterings} multiple electron-hole scatterings  as plotted in Fig.~\ref{fig:BS-eh} for systems with the repulsive interaction. In order to keep the analytic control we resort only to two scattering channels for the direct vertex. The second channel to be selected must be able to suppress the spurious transition and it is the \index{Bethe-Salpeter equation} Bethe-Salpeter equation with \index{multiple electron-electron scatterings} multiple electron-electron scatterings of Fig.~\ref{fig:BS-ee}.   

In the next step we separate the critical from the noncritical fluctuations. We know that the irreducible vertex from the electron-hole channel remains finite and only the reducible vertex of this channel becomes singular. The full parquet equation for the singlet irreducible vertex $\Lambda_{\uparrow\downarrow}$ contains a convolution of two singular reducible vertices $\mathcal{K}_{\uparrow\downarrow}$, which is the reason why the full set of \index{parquet equations} parquet equations miss completely the critical region of the RPA pole and the Kondo regime in the SIAM. Since we know that the Kondo effect is exact in the SIAM we assume that the superdivergent term, the convolution of two divergent terms, is compensated by higher-order contributions beyond the parquet equations with the bare interaction. With this assumption we can reduce the parquet equation for the irreducible vertex from the electron-hole channel to      
\begin{figure}
  \begin{center}
    \includegraphics[width=11cm]{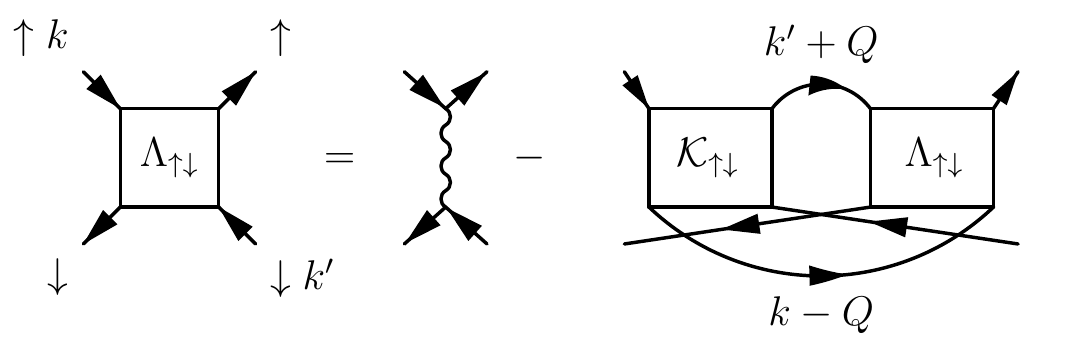}
   \caption{The reduced parquet equation for the irreducible vertex from the electron-hole channel in which the convolution of the divergent vertices $\mathcal{K}_{\uparrow\downarrow}$ is assumed to be compensated by higher order terms beyond the parquet equations.}  \label{fig:RPE-EE} 
  \end{center}  
 \end{figure}
\begin{multline}\label{eq:Lambda-reduced-bar}
 \Lambda_{\sigma\bar{\sigma}}(\veck,i\omega_{n};\veck',  i\omega_{n'}) = U 
 - \frac 1N\sum_{\vecQ}  \frac 1\beta\sum_{\nu_{l}} K_{\sigma\bar{\sigma}}(\veck,i\omega_{n}, \veck' + \vecQ,  i\omega_{n'} + i\nu_{l}; -\vecQ, - i\nu_{l} )
 \\
 \times  G_{\sigma}( \veck' + \vecQ,  i\omega_{n'} + i\nu_{l}) G_{\bar{\sigma}} (\veck -\vecQ , i\omega_{n} - i\nu_{l})  \Lambda_{\sigma\bar{\sigma}}(\veck' + \vecQ ,i \omega_{n'} + i\nu_{l },\veck - \vecQ, i\omega_{n} -  i\nu_{l})\,,
 \end{multline} 
the diagrammatic representation of which is plotted in Fig.~\ref{fig:RPE-EE}. The reducible vertex $\mathcal{K}_{\uparrow\downarrow}$ is determined from the unrestricted Bethe-Salpeter equation in the electron-hole channel. It reads
  \begin{multline}\label{eq:K-reduced-bar}
 K_{\sigma\bar{\sigma}}(\veck,i\omega_{n},\veck',i\omega_{n'}; \vecq, i\nu_{m}) 
 = - \frac 1N\sum_{\veck''}\frac 1\beta\sum_{\omega_{l}} \Lambda_{\sigma\bar{\sigma}}(\veck,i\omega_{n};\vecq + \veck'',  i\omega_{m + l}) G_{\bar{\sigma}} (\veck'' + \vecq, i \omega_{m + l}) 
 \\
 \times   G_{\sigma}(\veck'',i\omega_{l})\left[\Lambda_{\sigma\bar{\sigma}}(\veck'',i\omega_{l};\vecq + \veck',i\omega_{m + n'})
 %\right. \\ \left.
 +  K_{\sigma\bar{\sigma}}(\veck'',i\omega_{l},\veck', i\omega_{n'};\vecq,  i\nu_{m})  \right] 
 \end{multline}
\begin{figure}
  \begin{center}
    \includegraphics[width=15cm]{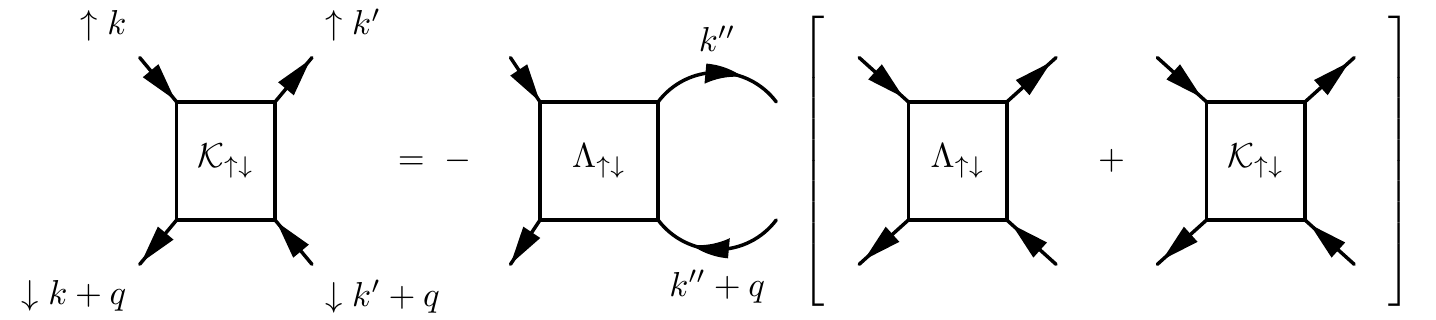}
   \caption{Bethe-Salpeter equation for the reducible vertex from the singular, electron-hole, scattering channel. The diagrams within the brackets are attached to the two-particle propagator in front of the brackets.}  \label{fig:RPE-EH} 
  \end{center}  
 \end{figure}
and is plotted in Fig.~\ref{fig:RPE-EH}.

Equations~\eqref{eq:Lambda-reduced-bar} and~\eqref{eq:K-reduced-bar} are the simplest parquet-like equations introducing a two-particle self-consistency without missing the weak-coupling critical behavior. They are solvable only numerically in their full generality. We still need a simplification in order not to lose the analytic  control of the approximation. We use the fact that we are essentially interested in the critical behavior in the proximity of the \index{RPA pole} RPA pole. We hence divide the dynamical fluctuations of the vertex functions into relevant ones, controlling the critical behavior, and  irrelevant ones, that do not affect the universal critical  behavior. We further neglect the irrelevant frequency and momentum dependence of the noncritical functions. The irrelevant variables are the fermionic ones. If we neglect them in the irreducible vertex we reduce it to a static effective interaction $\Lambda$. Using the solution for the reducible vertex $\mathcal{K}(\vecq,i\nu_{m})$ from Eq.~\eqref{eq:K-reduced-bar} in Eq.~\eqref{eq:Lambda-reduced-bar} we obtain   
\be\label{eq:Lambda-full}
\left[1 - \frac {\Lambda^{2}}N\sum_{\vecq}\frac 1\beta\sum_{\nu_{m}} \phi(-\vecq,-i\nu_{m})\frac{G_{\uparrow}(\veck + \vecq,i\omega_{n + m}) G_{\downarrow}(\veck'- \vecq,i\omega_{n'- m})}{1 + \Lambda\phi(-\vecq,-i\nu_{m})}  \right] \Lambda = U \,,
\ee
where
\be
\phi(\vecq,i\nu_{m}) = \frac 1{2N} \sum_{\sigma}\sum_{\veck}\frac 1\beta\sum_{\omega_{n}} \left[G_{\bar{\sigma}}(\veck + \vecq,i\omega_{n + m}) + G_{\bar{\sigma}}(\veck - \vecq,i\omega_{n - m})\right]G_{\sigma}(\veck,i\omega_{n}) \,.
\ee

It is evident that this equation cannot hold point-wise for all fermionic momenta and frequencies. It means that we can satisfy Eq.~\eqref{eq:Lambda-full} only approximately unless  we decorate the effective interaction $\Lambda$ with frequencies and momenta.  The dynamical vertex $\Lambda$ would make the resulting renormalization extremely complicated and would lead to losing the analytic control of the critical behavior, which we do  not want. If we cannot obey Eq.~\eqref{eq:Lambda-full} fully we resort to an approximate solution. There are two options how to close the equation for the effective interaction $\Lambda$. The first one is suitable for metallic systems at very low temperatures. It uses analytic continuation to real frequencies where we put the fermionic frequencies on the Fermi level. This approximation works well in the strong-coupling limit of the SIAM at zero temperature \cite{Janis:2017ab,Janis:2020aa}. The other option is to use the fact that the universal critical behavior is not qualitatively affected by fluctuations in the fermionic variables since they are irrelevant in the critical region of the pole in the response function.  We can then average Eq.~\eqref{eq:Lambda-full} over the fermionic momenta and Matsubara frequencies. The selection of the averaging procedure is not, however, unambiguous \cite{Janis:2008ab}. One way, particularly suitable for extended lattice systems, is to multiply both sides of Eq.~\eqref{eq:Lambda-full} by the product of the one-particle thermodynamic propagators $G_{\downarrow}(-\veck,-i\omega_{n})G_{\uparrow}(-\veck',-i\omega_{n'})$ and sum/integrate over the fermionic variables $\veck,\omega_{n}$ $\veck',\omega_{n'}$. Equation~\eqref{eq:Lambda-full} reduces after this averaging to  \cite{Janis:2020ac}    
\begin{equation}\label{eq:Lambda-phi2a}
\Lambda = \frac{U\left(n^{2} - m^{2}\right)}{n^{2} - m^{2} + 4\Lambda^{2} \mathcal{X} } \,,
\end{equation} 
where
\begin{subequations}\label{eq:X-psi2-phi}
\begin{align}
\mathcal{X} &= -\frac 1{N}\sum_{\vecq} \frac{\psi(\vecq,i\nu_{m})\psi(-\vecq,-i\nu_{m})\phi(-\vecq,-i\nu_{m})}{1 + \Lambda\phi(-\vecq, -i\nu_{m})} \,, 
\\
\psi(\vecq,\omega_{+}) &=  \frac 1{N} \sum_{\veck}\frac 1\beta\sum_{\omega_{n}} G_{\bar{\sigma}}(\vecq - \veck,i\omega_{m - n})G_{\sigma}(\veck,i\omega_{n}) \,.
\end{align}
\end{subequations}
We denoted the charge and spin densities $n= (\beta N)^{-1}\sum_{\sigma}\sum_{\veck,\omega_{n}}G_{\sigma}(\veck,i\omega_{n})e^{i\omega_{n}0^{+}}$ and $m=(\beta N)^{-1}\sum_{\sigma}\sigma\sum_{\veck,\omega_{n}}G_{\sigma}(\veck,i\omega_{n})e^{i\omega_{n}0^{+}}$, respectively. Positivity of integral $\mathcal{X}$ leads to a screening of the bare interaction. Equation~\eqref{eq:Lambda-phi2a}  self-consistently  determines the \index{effective interaction} effective interaction $\Lambda$ that controls the critical behavior near the pole, a singularity in the integrand of Eq.~\eqref{eq:X-psi2-phi}. The singularity emerges  when $a(\vecQ) = 1 + \Lambda\phi(\vecQ,0) =0$. Vector $\vecQ$ is the momentum at which the static bubble $\phi(\vecq,0)$ has a minimum, maximum of its modulus. It characterizes the type of the critical behavior we investigate.  The critical point can be reached and crossed to an ordered phase only if $\mathcal{X}<\infty$. It means that the physical singularity must be integrable.  The self-consistent equation for the effective interaction then suppresses the spurious poles of the random-phase approximation with the bare interaction $U$ and allows only for integrable singularities in the response functions.

\subsection{Spectral function}

Equations~\eqref{eq:Lambda-phi2a} and~\eqref{eq:X-psi2-phi} determine fully the effective interaction from which we derive most of the thermodynamic properties of the model. All calculations can be performed in the Matsubara formalism without the necessity to use analytic continuation to real frequencies. We showed that this  thermodynamic mean-field approximation suppresses the spurious phase transition of the Hartree approximation. It is the desired extension of the Hartree weak-coupling mean-field theory to the strong-coupling regime.  This thermodynamic theory remains essentially at the two-particle level. The one-particle propagators are an input and can be approximated via a separate scheme, except for the \index{odd self-energy} odd self-energy that reduces to a static value $\Delta\Sigma = -\Lambda m/2$. The \index{even self-energy} even self-energy is  obtained from an approximation to the corresponding Schwinger-Dyson equation.  

The Schwinger-Dyson dynamical equation for the static effective interaction reads 
 \begin{subequations}\label{eq:SDE-full}
 \begin{multline}\label{eq:SDE-Sigma}
 \bar{\Sigma}(\veck,\omega_{+}) =   U \frac n2 - \frac{U \Lambda}{N}\sum_{\vecq}P\int_{-\infty}^{\infty} \frac{dx}{\pi} \left\{b(x) \bar{\mathcal{G}}(\veck + \vecq, \omega_{+} + x) \Im\left[\frac{\bar{\Phi}(\vecq,x_{+})}{1 + \Lambda \phi(\vecq,x_{+})} \right]     
 \right. \\ \left.
 -\ \frac{f(x + \omega)\bar{\Phi}(\vecq, x_{-})}{1 + \Lambda\phi(\vecq, x_{-}) }\Im\bar{\mathcal{G}}(\vecq + \veck,x + \omega_{+})\right\} \,,
 \end{multline}
where we introduced a two-particle bubble with the renormalized one-particle propagators
\bdm\label{eq:SDE-bubble}
\bar{\Phi}(\vecq,\omega_{+}) = -\frac{1}{N}\sum_{\veck}\int_{\infty}^{\infty}\frac{dx}{\pi}f(x)\left[\bar{\mathcal{G}}(\veck + \vecq,x + \omega_{+}) + \bar{\mathcal{G}}(\veck - \vecq,x - \omega_{+}) \right]\Im \bar{\mathcal{G}}(\veck,x_{+}) \,.
\edm
 \end{subequations}
We used new symbols for the one-particle Green function $\bar{\mathcal{G }}$  and for the two-particle bubble  $\bar{\Phi}$ on the right-hand side of Eq.~\eqref{eq:SDE-full}. Independently of which even self-energy we used in the \index{thermodynamic Green functions} thermodynamic Green functions $G_{\sigma}(\veck,i\omega_{n})$ in Eqs~\eqref{eq:Lambda-phi2a} and~\eqref{eq:X-psi2-phi} to determine $\Lambda$ and $\phi(\vecq,\omega_{+})$, the Green function $\bar{\mathcal{G}}(\vecq,\omega_{+})$ must be  at least partly renormalized by the resulting self-energy $\bar{\Sigma}(\veck,\omega_{+})$ from Eq.~\eqref{eq:SDE-full}. The simplest renormalization would be the \index{Hartree self-consistency} Hartree self-consistency where the self-energy in the one-particle propagator $\bar{\mathcal{G }}$ on the right-hand side of Eq.~\eqref{eq:SDE-full} is approximated by the linear term in the interaction strength. That is
\begin{align}\label{eq:Hartree-SC}
\bar{\Sigma}(\veck,\omega_{+}) &=   U \frac n2 = - \frac 1N\sum_{\veck}\int_{-\infty}^{\infty} \frac{dx}{\pi}f(x)\Im \bar{\mathcal{G}}(\veck,x_{+}) \,
\end{align}
is used in the Green function $\bar{\mathcal{G}}$ in Eq.~\eqref{eq:SDE-full}. It is the mean-field selection of the even self-energy in the two-particle thermodynamic calculations and it closes the static mean-field approximation with a two-particle self-consistency.  

The importance of the \index{Schwinger-Dyson equation} Schwinger-Dyson equation lies in its dynamical structure that is affected in the critical region by the singular two-particle propagator. If we go beyond the static Hartree approximation for $\bar{\Sigma}(\veck,\omega_{+}) $ the full one-particle self-consistency in Eq.~\eqref{eq:SDE-full} demands that we use the following fully  renormalized one-particle Green function in models with a magnetic critical behavior  
\begin{align}\label{eq:G-renorm}
\mathcal{G}_{\sigma}(\veck,  i\omega_{n}) &= \frac 1{i\omega_{n} + \mu - \sigma\Delta\Sigma  -\epsilon(\veck) - \bar{\Sigma}(\veck,i\omega_{n})} \,.
\end{align}

Advantage of the two-particle approach is the possibility to separate the one-particle  self-consistency from the two-particle one. It means that the one-particle Green functions $G_{\sigma}$ determining the effective interaction and the two-particle bubble can differ from the Green function $\bar{\mathcal{G}}$ in Eq.~\eqref{eq:G-renorm}. We demonstrated that the best approximation of the Kondo limit and the three-peak structure of the spectral function of the SIAM deliver the thermodynamic propagators with only the static Hartree renormalization \cite{Janis:2017aa}. 
%It appears, however, that the static even self-energy from Eq.~\eqref{eq:Hartree-SC}  should be determined from the Green function $\mathcal{G}$ from Eq.~\eqref{eq:G-renorm}. Otherwise the approximation may lead to negative compressibility \cite{Janis:2017ab}.       

If we renormalize the one-particle Green functions in the parquet equations differently from the Schwinger-Dyson equation then the thermodynamic propagators in the two-particle approach play the role of  the bare propagators from the unrenormalized perturbation theory. The physical self-energy and the Green function are those determined from Eqs.~\eqref{eq:SDE-full} and~\eqref{eq:G-renorm}. An important restriction on the even self-energy from the Schwinger-Dyson equation is that it depends only on even powers of the symmetry-breaking field and hence, does not affect the critical behavior derived from the two-particle irreducible vertex $\Lambda$ from the critical Bethe-Salpeter equation.

\section{Conclusions}

It is generally difficult to calculate thermodynamic properties of the interacting many-body systems even for simplified models. The exact solutions are rare and available only in specific limiting situations. Understanding the impact of the particle interaction on the collective macroscopic behavior is of great importance for our ability to control and utilize material properties to for our convenience. Methods of classical statistical physics are sufficient to reliably describe macroscopic phenomena at high, room temperatures. Quantum physics comes into play at lower temperatures where quantum many-body models must be used to explain the long-range effects in solids. 

Microscopic quantum dynamics, noncommutativity of the fundamental operators, and indistinguishability of the identical particles makes it extremely difficult to get quantitative results when the interaction strength is dominant over the kinetic  energy. It is one of the most important tasks of the condensed-matter theory to develop reliable and generally applicable techniques in the strong-coupling regime of quantum many-body models. The quantum perturbation theory with Feynman diagrams and many-body Green functions becomes one of the most powerful means to reach the goal.

The only way to use the perturbation theory in strong coupling is to use renormalizations and self-consistent summations of classes of Feynman diagrams. This can be done only by using many-body Green functions and their analytic properties. Renormalizations of the perturbation theory must be done carefully and in a controlled way not to lose causality and/or not to break conservation laws. The major problem of quantum interacting models is the two-fold interconnectivity of one and two-particle functions. The first one is the dynamical equation of motion, being the Schwinger-Dyson equation for the Green functions. The other is the Ward identity.  

The standard way to introduce renormalizations into the many-body perturbation theory is the Baym-Kadanoff construction with the Luttinger-Ward functional generating the self-energy as the fundamental object of this approach. It formally guarantees the conserving character of the theory but no approximate self-energy from the Schwinger-Dyson equation obeys the Ward identity. There are two two-particle vertices to the approximate self-energy. That is why this approach cannot be continued beyond the critical points to the ordered phases.

An alternative way to introduce renormalizations into the perturbation theory is to build the theory from the two-particle vertex. After identifying the critical behavior of the model, one chooses the irreducible vertex of the singular Bethe-Salpeter equation as the generating functional. One splits the Green functions into either odd or even symmetric parts with respect to the reflection of the symmetry-breaking field responsible for the critical behavior.  Only the two-particle functions with even symmetry  become relevant when we use the Ward identity valid up to the linear order of the symmetry-breaking field. The odd part of the one-particle self-energy is related with the two-particle irreducible vertex via a \index{Ward identity-linearized} linearized Ward identity. The even self-energy then obeys the symmetrized Schwinger-Dyson equation. The separation of the odd and even self-energies seems to be  the only way to reconcile at least qualitatively the Ward identity and the Schwinger-Dyson equation in approximate theories.    
     
The two-particle approach is from obvious reasons much more complicated than the one-particle one of Baym and Kadanoff. We nevertheless succeeded to reduce the complexity of the two-particle functions by neglecting noncritical fluctuations and keeping only the critical ones. In this way we were able to construct a mean-field theory with a static effective interaction being a consistent extension of the RPA to strong coupling. It naturally includes a two-particle self-consistency in the effective interaction suppressing thereby the spurious pole of the RPA. It separately allows for various levels of renormalizations of the even dynamical self-energy in the Schwinger-Dyson equation. Last but not least, the two-particle mean-field approximation is consistent with the Mermin-Wagner theorem and distinguishes between zero and non-zero temperatures in low spatial dimensions.

\section*{Acknowledgment}
I thank Antonín Klíč, Vladislav Pokorný, Ania Kauch, and Jiawei Yan for their contributions to the development  of the two-particle approach and the application of the reduced parquet equations in impurity models.  I acknowledge support from Grant No.~19-13525S of the Czech Science Foundation.   

%%%%%%%%%%%%%%% Appendices %%%%%%%%%%%%%%%%%
%\newpage % if you want the appendix to start on a new page
%\section*{Appendices}
%\appendix
%\section{Some remarks}

%%%%%%%%%%%%%%% References %%%%%%%%%%%%%%%%%
\clearpage
%% or, should you use bibtex:
%\bibliographystyle{correl}
%\bibliography{../../BibTeX/parquets_MB,../../BibTeX/Impurity_solver,../../BibTeX/Mean_Field_Approx,../../BibTeX/CPA,../../BibTeX/Green_functions,../../BibTeX/1D-Hubbard1}
%

\clearchapter

% To generate the index run {\tt makeindex} on your manuscript.
% You should mark about 20 important concepts that are explained in your text,
%  which will be collected in the index. \verb|\index{indexEntry}| has to be added
% in the text next to the word to which the indexEntry refers; ''indexEntry'' is
% the key-word that will appear in the index. Example:
%	\begin{verbatim}
%	In the last two decades, we have witnessed breathtaking advances 
%	in the realistic modeling \index{realistic modeling} of materials 
%	with strong correlations. \index{strongly correlated materials}
%	\end{verbatim}
%	Make sure that index entries only refer to pages that actually discuss that concept.
%	Please, use only lower-case in indexEntry, except for names like, e.g., in 'Hubbard model'.
%	If you have several related concepts, you can group them, e.g., 
%	\begin{verbatim}
%	 \index{correlation}
%	 \index{correlation!electronic}
%	 \index{correlation!effects}
%	\end{verbatim}

\end{document}